\documentclass[11pt,a4paper]{article}
\pdfoutput=1

\usepackage{mathtools, nccmath}
\usepackage{verbatim}
\usepackage{amsfonts}
\usepackage{amssymb}
\usepackage{amsmath}
\usepackage{braket}
\usepackage{dsfont}
\usepackage{authblk}
\usepackage[margin=1.3cm]{caption}
\usepackage{blkarray}
\usepackage{graphicx}
\usepackage{caption}
\usepackage{subcaption}
\usepackage{slashed}
\usepackage{color}
\usepackage{bm}

\usepackage{jheppub}
\date{}

\numberwithin{equation}{section}

\newcommand{\mL}{\mathcal{L}}
\newcommand{\mO}{\mathcal{O}}
\newcommand{\mJ}{\mathcal{J}}

\newcommand{\pd}{\partial}

\begin{document}

\author{Gabriel Cuomo}
\affiliation{Theoretical Particle Physics Laboratory (LPTP), Institute of Physics, EPFL, Lausanne, Switzerland}
\emailAdd{gabriel.cuomo@epfl.ch}

\abstract{We include vortices in the superfluid EFT for four dimensional CFTs at large global charge. Using the state-operator correspondence, vortices are mapped to charged operators with large spin and we compute their scaling dimensions. Different regimes are identified: phonons, vortex rings, Kelvin waves, and vortex crystals. We also compute correlators with a Noether current insertion in between vortex states. Results for the scaling dimensions of traceless symmetric operators are given in arbitrary spacetime dimensions.}

\title{Superfluids, vortices and spinning charged operators in 4d CFT}

\keywords{Conformal field theory, Effective Field Theories, Global Symmetries.}

\arxivnumber{1906.07283}

\maketitle

\date{}

%
%
%

%

\section{Introduction}

Conformal field theories (CFTs) play a key role in particle and condensed matter physics.
As fixed points of the renormalization group flow, they act as landmarks in the space of quantum field theories (QFTs). Through the AdS/CFT correspondence \cite{Maldacena,Witten}, they promise to shed light on quantum gravity. They also describe critical points for second order phase transitions. Finally, CFTs are also among the few examples of interacting QFTs where exact results are available without supersymmetry. Recently, the bootstrap program \cite{Polyakov,Rattazzi} achieved much progress in the study of CFTs, both through numerical 
\cite{BootstrapIsing,Vichi_review} and analytical
\cite{LargeS1,LargeS2,InversionFormula1} techniques.

Basic observables in CFTs are correlation functions of local operators in the vacuum. Despite this, sometimes one can make predictions for the CFT data defining the theory studying the dynamics of finite density states \cite{Hellerman}. This is a consequence of the state/operator correspondence \cite{Rychkov_lectures,SD_lectures}, which relates states in radial quantization to local operators with the same quantum numbers. 

This idea has been applied in the investigation of the superfluid phase in conformal field theories
\cite{Hellerman,MoninCFT,Bern1,Bern2,Bern3,Bern4,Bern5,Bern6,CSLargeQ}. Indeed superfluids
are the most natural candidates for the description of states at large internal quantum numbers in CFTs. They admit a simple and universal effective field theory (EFT) description \cite{SonSuperfluid,Nicolis_Zoology} which allows the computation of correlators in a perturbative expansion controlled by the charge density.
Recently, the same strategy was applied in the context of non-relativistic CFTs \cite{NRCFTLargeQ1,NRCFTLargeQ2,NRCFTLargeQ3}.

As the angular momentum is increased, the superfluid starts rotating 
and vortices develop
\cite{Vortex0}. These can be included in the EFT as heavy topological defects \cite{Lund,Davis,Vortices,Vortices1,Vortices2}. In \cite{Cuomo}, this EFT was used to describe operators with large spin and large charge in three dimensional CFTs. In this work, we study the predictions of the vortex EFT for four dimensional CFTs.

\subsection{Summary of results}

Let us first set our conventions for the four dimensional rotation group $SO(4)$. Spinning operators in four dimensions are classified in representations labelled by two positive half-integer quantum numbers $(J,\bar{J})$. These are related to the maximal values allowed for the Cartan generators $J_{34}$ and $J_{12}$ as
\begin{equation}\label{eqSummaryJJ}
(J,\bar{J})=\left(
\frac{|J_{34}-J_{12}|}{2},
\frac{|J_{12}+J_{34}|}{2}
\right).
\end{equation}
With no loss of generality, we assume $J_{34}\geq J_{12}\geq 0$.

Consider a CFT invariant under an internal $U(1)$ symmetry.
The main prediction of the superfluid EFT is the scaling dimension of the lightest scalar operator of charge $Q$ in the spectrum. It is given by \cite{MoninCFT}
\begin{equation}\label{eqSummaryDelta0}
\Delta_0(Q)=\alpha Q^{4/3}+\beta Q^{2/3}+\ldots,
\end{equation}
for $Q\gg 1$; here $\alpha$ and $\beta$ are independent Wilson coefficients. 

In this work, we compute the scaling dimension of the lightest operator as the spin is increased.
As in \cite{Cuomo}, the EFT describes the regime where the spin is below the unitarity bound, $J,\bar{J}\ll Q^{4/3}$, and cannot reach the regime analyzed by the analytic bootstrap \cite{LargeS1,LargeS2,LargeS_Mat1,LargeS_Alday1,LargeS_Alday2,
LargeS_Alday3,LargeS_Alday5,LargeS_Alday6,
LargeS_Kaviraj1,LargeS_Kaviraj2,LargeS_Ising}. To leading order in the charge and the spin, the results depend on the first coefficient in \eqref{eqSummaryDelta0} and on an extra Wilson coefficient $\tilde{\gamma}$ parametrizing the vortex tension. 

For traceless symmetric operators $J=\bar{J}=J_{34}/2$, the corresponding state passes through three distinct regimes, qualitatively similar to the CFT$_3$ case:
\begin{itemize}
\item For $2\leq J_{34}\ll Q^{1/3}$ the lightest operator corresponds to a phonon wave of angular momentum $J$ in the superfluid. The scaling dimension is given by
\begin{equation}\label{eqResult1Phonon}
\Delta=\Delta_0(Q)+\sqrt{\frac{J_{34}(J_{34}+2)}{3}}+\mO\left(\frac{J_{34}^4}{Q^{2/3}}\right).
\end{equation}
\item For $Q^{1/3}\ll J_{34}\leq Q$, the minimal energy state is given by a single vortex ring, whose radius increases with $J$. Its energy is
\begin{equation}\label{eqResult1Vortex}
\Delta=\Delta_0(Q)+\Delta_V(Q,J_{34}),
\end{equation}
where
\begin{multline}\label{eqSummaryDelta1Vortex}
\Delta_V(Q,J)\equiv\frac{3}{8\alpha}Q^{1/6}J^{1/2} 
\log \left(  J/Q^{1/3}\right)
-\frac{3}{4\alpha}Q^{1/6}J^{1/2}\log \left(1+\sqrt{J/Q}\right)\\
-\frac{3}{2\alpha}Q^{2/3}\log \left(1+\sqrt{J/Q}\right)
+\tilde{\gamma}Q^{1/6}J^{1/2}+
\mO\left(Q^{1/6}J^{1/2} \times \frac{Q^{1/3}}{J}\right).
\end{multline}
The leading contribution in \eqref{eqSummaryDelta1Vortex} comes from the first term, because of the logarithmic enhancement. The other terms can be interpreted as finite-size corrections due to the vortex extension and are functionally distinguished from the relative $Q^{1/3}/J$ corrections.
\item For $Q\ll J_{34}\ll Q^{4/3}$ the superfluid forms a vortex crystal. The scaling dimension of the corresponding operator is given by 
\begin{equation}\label{eqResult1VortexCrystal}
\Delta=\Delta_0(Q)+\frac{3}{4\alpha}\frac{J_{34}^2}{Q^{4/3}}+
\mO\left(\frac{J_{34}^2}{Q^{4/3}}\times\frac{Q}{J_{34}},\frac{J_{34}^2}{Q^{4/3}}\times
\left(\frac{J_{34}}{Q^{4/3}}\right)^2\right).
\end{equation}
\end{itemize}
Mixed symmetric representations are conveniently parametrized in terms of $J_{34},J_{12}$ in \eqref{eqSummaryJJ}. We write $J_{ab}$ to generically
denote any of them.
We find the following results:
\begin{itemize}
\item For $2\leq J_{12}\leq J_{34}\ll Q^{1/3}$ the minimal energy state is given by two phonons propagating on the superfluid, with energy:
\begin{equation}\label{eqResult2Phonons}
\Delta=\alpha Q^{4/3}+\sqrt{\frac{J_{34}(J_{34}+2)}{3}}+\sqrt{\frac{J_{12}(J_{12}+2)}{3}}+
\mO\left(\frac{J_{ab}^4}{Q^{2/3}}\right).
\end{equation}
\item For $1\leq Q-J_{34}\ll Q$ and $2\leq J_{12}\ll Q^{1/3}$, the lowest energy state corresponds to a Kelvin wave of spin $J_{12}$ propagating on a large vortex ring. The corresponding operator scaling dimension is given by:
\begin{multline}\label{eqResult1Kelvon}
\Delta=\alpha Q^{4/3}+
\Delta_{V}(Q,J_{34})\\
+\frac{3}{8\alpha}\frac{\pi(J_{12}^2-1)}{Q^{1/3}}
\left[\log Q^{2/3}-2 \psi\left(\frac{J_{12}+1}{2}\right)-2\gamma_E-1-\log 64\right]\\
+\tilde{\gamma}\frac{\pi(J_{12}^2-1)}{Q^{1/3}}
+\mO\left(\frac{J_{12}^4}{Q}\right).
\end{multline}
\item For $Q^{1/3}\ll J_{12}\leq J_{34}\leq Q$ and $\left(J_{12}+J_{34}-Q\right)^2\gg J_{12}J_{34}/Q^{2/3}$, the minimal energy state is given by two vortex rings. When $1\leq Q-J_{34}\ll Q^{1/3}$ the energy is given by the sum of the two free contributions
\begin{equation}\label{eqResult2Vortices}
\Delta=
\alpha Q^{4/3}+
\Delta_{V}(Q,J_{34})+\Delta_{V}(Q,J_{12}), \qquad 1\leq Q-J_{34}\ll Q^{1/3}.
\end{equation}
Interactions correct the result in the general case, which takes the same form only to logarithmic accuracy
\begin{equation}\label{eqResult2VorticesGuess}
\Delta=\alpha Q^{4/3}+\frac{3}{8\alpha}Q^{1/6}\left[J_{34}^{1/2}\log \left(  J_{34}/Q^{1/3}\right)
+J_{12}^{1/2}\log \left(  J_{12}/Q^{1/3}\right)\right]
+\mO\left(Q^{1/6}J_{ab}^{1/2}\right).
\end{equation}
\item For $Q\ll J_{12}\leq  J_{34}\ll Q^{4/3}$ the superfluid arranges in a vortex lattice as in \eqref{eqResult1VortexCrystal}; the scaling dimension of the corresponding operator is
\begin{equation}\label{eqResult2VortexCrystal}
\Delta=\alpha Q^{4/3}+\frac{3}{4\alpha}\frac{J_{34}^2+J_{12}^2}{Q^{4/3}}+
\mO\left(\frac{J_{ab}^2}{Q^{4/3}}\times\frac{Q}{J_{ab}},\frac{J_{ab}^2}{Q^{4/3}}\times
\left(\frac{J_{ab}}{Q^{4/3}}\right)^2\right).
\end{equation}
\end{itemize}
These results apply to CFTs whose large charge sector can be described as a superfluid and which admit vortices. These are natural and simple conditions, hence we expect them to apply to a wide range of theories with a $U(1)$ symmetry. Nonetheless, we cannot prove these assumptions.

The rest of the paper is organized as follows. In section \ref{secReview}
we review the superfluid description of large charge operators as well as the vortex EFT in $2+1$ dimensions. In \ref{secEFTformulation} we 
formulate the effective field theory (EFT) for vortices in $3+1$ dimensions. The results presented above are derived in sec. \ref{secVortexEFT}. In \ref{secCorrelators} we show how to make predictions for correlators involving a current insertion between two vortex states and in \ref{secGeneralD} we briefly comment on how the results \eqref{eqResult1Vortex} and \eqref{eqResult1VortexCrystal} change in generic spacetime dimensions. Finally in \ref{secConclusions} we draw our conclusions and comment on future research directions. Technical details are given in the appendices \ref{AppProp}, \ref{AppMagnetostaticEnergy}, \ref{AppendixCoset}.

\vspace{0.2cm}
\emph{Conventions and coordinates on $S^3$:} Lorentz indices $\mu,\nu,\ldots$ go from $0$ to $3$ and we use mostly minus metric signature $\text{sgn}(g_{\mu\nu})=\{1,-1,-1,-1\}$. Spatial indices are written as $i,j,\ldots=1,2,3$ and are raised and lowered with a positive metric $|g_{ij}|$. We use the notation $\dot{f}=\pd_0f$ for time derivatives. Indices $a,b,\ldots$ are used for the $\mathds{R}^4$ embedding of $S^3$ and go from $1$ to $4$. Embedding coordinates are denoted $X_a=X^a$. Calling $X_a(x)$ the $\mathds{R}^4$ coordinate corresponding to an $S^3$ point $x$, the chordal distance between two points $x$ and $x'$ is given by:
\begin{equation}
\Delta X^2(x,x')=\sum_a\left[X_a(x)-X_a(x')\right]^2.
\end{equation}
A convenient parametrization of $S^3$ is provided by Hopf coordinates, defined via the embedding:
\begin{equation}\label{eqHopf}
X_1=R\cos \xi\sin\eta,\quad X_2=R\sin \xi\sin\eta,\quad 
X_3=R\cos\phi\cos\eta,\quad
X_4=R\sin\phi\cos\eta.
\end{equation}
This gives the following metric tensor
\begin{equation}\label{eqHopf2}
\frac{ds^2}{R^2}=d\eta^2+\sin^2\eta d\xi^2+\cos^2\eta d\phi^2,\qquad
\eta\in[0,\pi/2],\quad\xi\in [0,2\pi],\quad\phi\in [0,2\pi].
\end{equation}
For fixed $\eta$ different from $0$ and $\pi/2$, $\xi$ and $\phi$ describe an $S^1\times S^1$ submanifold.

\section{Review of previous results}\label{secReview}

\subsection{Conformal superfluid}

Let us first remind that, in a CFT, the state-operator correspondence relates eigenstates of the Hamiltonian $H$ on $S^d$ with the set of local operators at any given point \cite{Rychkov_lectures,SD_lectures}. The quantum numbers of the state on $S^d$ and the corresponding operator are the same. In particular, the energy $E$ is related to the scaling dimension of the latter as $\Delta=E/R$.

The EFT description of CFTs at large quantum numbers is based on the assumption that the lightest scalar operator with $U(1)$ charge $Q$ in a $d+1$ dimensional CFT corresponds to a state with homogeneous charge density on $\mathds{R}\times S^d$. For $Q\gg 1$, the scale associated with the density of this state is parametrically bigger than the $S^d$ radius $R$ and the CFT is expected to be in a ``condensed matter phase''. As argued in \cite{MoninCFT}, the simplest possibility is that the CFT enters a superfluid phase. Technically, this is equivalent to assuming an effective description in terms of a $U(1)$ Goldstone boson \cite{SonSuperfluid}. The effective Lagrangian is fixed by shift symmetry and Weyl invariance:
\begin{equation}\label{eqActionSuperfluid}
\mL/\sqrt{g}=c(\pd\chi)^{d+1}+c_1(\pd\chi)^{d-1}\left\{\mathcal{R}
+d(d-1)\frac{\left[\pd_\mu(\pd\chi)\right]^2}{(\pd\chi)^2}\right\}
+c_2(\pd\chi)^{d-1}\mathcal{R}_{\mu\nu}\frac{\pd^\mu\chi\pd^\nu\chi}{(\pd\chi)^2}
+\ldots.
\end{equation}
We use the notation $(\pd\chi)=\left(\pd_\mu\chi\pd^\mu\chi\right)^{1/2}$ and
$c,c_1,c_2$ are Wilson coefficients. Here $\mathcal{R}^\mu_{\;\nu\rho\sigma}$ is the Riemann tensor on the cylinder $\mathds{R}\times S^d$. We assume $c,c_1,c_2\sim\mO(1)$, corresponding to the generic expectation for a strongly coupled system. 
On a homogeneous background at finite charge, the field takes the value $\chi=\mu t$, where $\mu$ is the chemical potential of the system. To leading order in derivatives, it is related to the $U(1)$ charge density $j_0$ as
\begin{equation}\label{eqCurrentDdim}
j_0=\frac{Q}{R^d\Omega_d}=c(d+1)\pd_0\chi(\pd\chi)^{d-1}=(d+1)c\mu^d,
\end{equation}
where $\Omega_d=2\pi^{\frac{d+1}{2}}/\Gamma\left(\frac{d+1}{2}\right)$ is the $S^d$ volume.
The chemical potential sets the cutoff of the EFT:
\begin{equation}\label{eqDualCutoff}
\Lambda\sim \mu\sim\frac{Q^{1/d}}{R}.
\end{equation}
By the state/operator correspondence, the ground state of \eqref{eqActionSuperfluid} corresponds to the minimal energy state with charge $Q$. Its energy is determined by a semiclassical analysis and takes the form:
\begin{equation}\label{eqDeltaQdDimensions}
\Delta_0(Q)=\alpha Q^{\frac{d+1}{d}}\left(1+\frac{\beta}{\alpha}\, Q^{-\frac{2}{d}}
+\ldots\right).
\end{equation}
Quantum corrections provide a $Q^0$ contribution. For even $d$, there is no local counterterm correcting this term, which is hence universal \cite{Hellerman}.

The Lagrangian \eqref{eqActionSuperfluid} describes also excitations on the background.
For instance, expanding $\chi=\mu t+\pi$ and working to leading order in derivatives we get
\begin{equation}
\mL/\sqrt{g}=c\mu^{d-1}\frac{d(d+1)}{2}\left(\dot{\pi}^2
-\frac{1}{d}\pd_i\pi|g^{ij}|\pd_j\pi\right)+\ldots.
\end{equation}
Quantizing the system, it follows that the spectrum can be organized as a Fock space in terms of single particle states with angular momentum $J$ and energy given by
\begin{equation}\label{eqPhonon}
\omega_J=c_s\lambda_J,\qquad
c_s^2=\frac{1}{d}.
\end{equation}
Here $\lambda_J^2=\frac{J(J+d-1)}{R^2}$ are the eigenvalues of the Laplacian on $S^d$ and the sound speed $c_s$ is fixed by conformal invariance.
Physically, these states correspond to \emph{phonons} propagating in the superfluid and are associated with primary operators in the CFT. The $J=1$ mode has $\omega_1=1/R$ and corresponds to the creation of a descendant.

A natural question is how the spectrum changes as the spin $J$ is increased. When the angular momentum is parametrically smaller than the cutoff \eqref{eqDualCutoff}, the spectrum is reliably described by phonons \eqref{eqPhonon}. The results \eqref{eqResult1Phonon} and \eqref{eqResult2Phonons} then follow. Increasing spin, one finds singular solutions with a non zero winding number, such as $\pi=\phi$, where $\phi$ is the azimuthal angle. This signals that for $J\gg Q^{1/d}$ vortices develop in the superfluid and must be included in the effective description. This was done in \cite{Cuomo} for a $2+1$ dimensional CFT. The main goal of this work is to carry a similar analysis for a $3+1$ dimensional conformal field theory. To build some intuition, we briefly discuss the results of the $2+1$ dimensional EFT in the next section.

\subsection{Vortices in 2+1 dimensions}\label{secVortices3D}

In $d=2$, the action \eqref{eqActionSuperfluid} reads
\begin{equation}\label{eqAction3D}
\mathcal{L}=c(\pd\chi)^3.
\end{equation}
To study vortices, it is convenient to consider a dual description in terms of a gauge field. To this aim, we introduce an
independent variable $v_\mu \equiv\pd_\mu\chi$ and a Lagrange multiplier $A_{\mu}$ to set the curl of $v_\mu$ to zero:
\begin{equation}
\mathcal{L}=c v^3-\frac{1}{2\pi}A_\mu\frac{\epsilon^{\mu\nu\rho}}{\sqrt{g}}\pd_\nu v_\rho,
\end{equation}
where $\epsilon^{\mu\nu\rho}/\sqrt{g}$ is the antisymmetric Levi-Civita tensor.
Integrating out $v_\mu$ we get
\begin{equation}\label{eqDualGauge3d}
\mL=-\kappa F^{3/2}
\end{equation}
where $F=\sqrt{F_{\mu\nu}F^{\mu\nu}}$ and $F_{\mu\nu}=\pd_\mu A_\nu-\pd_\nu A_\mu$. The coefficient $\kappa$ is related to $c$ as $\kappa=\frac{1}{2^{5/4}(3\pi)^{3/2}\sqrt{c}}$.  
The $U(1)$ current relates the two descriptions
\begin{equation}
j^\mu=3c(\pd\chi)\pd_\mu\chi=\frac{1}{4\pi}\frac{\epsilon^{\mu\nu\lambda}}{\sqrt{g}}F_{\nu\lambda}.
\end{equation}
As a consequence, the charge density \eqref{eqCurrentDdim} translates into a homogeneous magnetic field $\langle F_{\theta\phi}\rangle=B\sin\theta=\frac{Q}{2R^2}\sin\theta$, which sets the cutoff of the EFT according to \eqref{eqDualCutoff}. 

The action \eqref{eqDualGauge3d} describes a propagating degree of freedom, given by the fluctuations of the magnetic field $F_{\theta\phi}$ and which corresponds to the phonon in the original picture, together with a non-propagating Coulomb field $A_0$, which does not have any local analogue in the scalar formulation. As we will see, it is precisely this extra component which provides the leading coupling to the vortices.

In the EFT, vortices are heavy charged particles in the dual description \eqref{eqDualGauge3d}. They are treated as $0+1$ dimensional worldlines, whose spacetime trajectory is parametrized by a function $X^\mu_p(\tau)$ of a time parameter $\tau$. The action of the superfluid plus vortices is fixed by the requirement of Weyl invariance and $\tau$-reparametrization invariance; the lowest orders in derivatives take the form \cite{Cuomo}
\begin{equation}\label{eqAction3dVortices}
S=-\kappa\int d^3x F^{3/2}-\sum_p q_p\int A_\mu dX^\mu_p-
\sum_p\int d\tau\sqrt{F}\sqrt{g_{\mu\nu}\dot{X}^\mu_p\dot{X}^\nu_p}
\,F_p\left(\frac{j_\mu \dot{X}^\mu}{j \dot{X}}\right).
\end{equation}
The second term is the minimal coupling between the gauge field and a particle of charge $q_p$; this cannot be written in a local form in the scalar picture, showing the convenience of the gauge formulation. Notice that the charge $q_p$ corresponds to the Goldstone winding number around $x_p$ and is hence quantized: $q_p\in\mathds{Z}$. The third term is the action for a relativistic point particle in a superfluid\footnote{See appendix \ref{AppendixCoset} for a derivation from the coset construction.}; it is multiplied by an arbitrary function of $\frac{j_\mu \dot{X}^\mu}{j \dot{X}}$, since the superfluid velocity breaks Lorentz symmetry and allows constructing an alternative \emph{condensed matter metric} \cite{Vortex0}. 

Working in the physical gauge $X_p^0=\tau$, we notice that the leading term in time derivatives for the vortex lines arises from the second piece in \eqref{eqAction3dVortices}. As we will self-consistently see, this implies that vortices move with non-relativistic velocities $|\dot{\vec{X}}|\sim1/\sqrt{B}$. Hence we can neglect terms with two time derivatives in the last term, retaining only a constant contribution proportional to $\sqrt{B}$ which is interpreted as the vortex mass. This procedure is sometimes called \emph{lowest Landau level} approximation in the literature \cite{integrateLandau1,integrateLandau2,integrateLandau3,jackiw1,jackiw2}.

The equations of motion (EOMs) deriving from \eqref{eqAction3dVortices} are
\begin{equation}\label{eqGaugeEOM3d}
\frac{1}{e^2}\nabla_i f^{ij}=\sum_p q_p\dot{X}_p^j\frac{\delta^2\left(x^i-X^i_p\right)}{\sqrt{g}},
\qquad
\frac{1}{e^2}\nabla_i E^i=\sum_p q_p\frac{\delta^2\left(x^i-X^i_p\right)}{\sqrt{g}},
\end{equation}
\begin{equation}\label{eqParticleEOM3d}
E^i=(\dot{X}_p)_jF^{ji},
\end{equation}
where $E^i=F^{i0}$ is the electric field and 
$e^2=\frac{2^{1/4}\sqrt{B}}{3\kappa}$. The particle EOMs \eqref{eqParticleEOM3d} are first order in derivatives and imply that vortices move with drift velocity $|\dot{\vec{X}}_p|\sim |\vec{E}/B|\sim 1/\sqrt{Q}$ as anticipated. Consequently, particle velocities, as well as the magnetic field fluctuations sourced by them, are negligible and the only relevant interaction is the electrostatic one.

We now look for static classical solutions of the EOMs \eqref{eqGaugeEOM3d} and \eqref{eqParticleEOM3d}. Because of the state/operator correspondence, classical solutions will be associated to operators with the same quantum numbers. The spin and the scaling dimension of the corresponding operators are then determined classically from the energy momentum tensor. The scaling dimension of a state with $n$ vortices reads\footnote{Here we correct a typo in eq. (19) of \cite{Cuomo}.}
\begin{multline}\label{eqDelta3d}
\Delta=\Delta_0(Q)+\frac{R}{2e^2}\int d^2x\sqrt{g}\vec{E}^2+\tilde{\gamma}\sum_p\sqrt{2}R\sqrt{B}\\
=
\Delta_0(Q)-\frac{\sqrt{Q}}{12\alpha}\sum_{p\neq r}q_pq_r
\log Q\Delta X^2(x_p,x_r)+\tilde{\gamma}n \sqrt{Q},
\end{multline}
where $X_p^a=\left(\sin\theta_p\cos\phi_p,\sin\theta_p\sin\phi_p,\cos\theta_p\right)$ is the vortex coordinate in the $\mathds{R}^3$ embedding of $S^2$ and 
$\Delta X^2(x_p,x_r)=\sum_{a=1}^3\left(X_p^a-X_r^a\right)^2$ is the chordal distance between two vortices. The first term in \eqref{eqDelta3d} is the energy of the homogeneous phase, given by \eqref{eqDeltaQdDimensions} with $d=2$:
\begin{equation}\label{eqDeltaZeroQ3d}
\Delta_0(Q)=\alpha Q^{3/2}+\beta Q^{1/2}+\ldots.
\end{equation} 
The second term is the energy stored in the electric field sourced by the vortices, which is further rewritten as a sum over pairwise contributions in the right-hand side; the $\log Q\sim\log\Lambda^2$ contribution arises from the logarithmically divergent self-energy of the point charges. We used
that the net charge on the sphere must be zero $\sum_p q_p=0$, as required by consistency of Gauss law on the sphere.
Finally, the last term is the contribution of $n$ vortex masses and is written in terms of an independent coefficient $\tilde{\gamma}$, assumed to be the same for all vortices. 

Similarly, the angular momentum is
\begin{equation}\label{eqAngularMom3d}
J_a=\frac{RB}{e^2}\int d^2x\sqrt{g}\,n^i_a\epsilon_{ij}\sqrt{g}E^j=-\frac{Q}{2}\sum_pq_p X_p^a.
\end{equation}
Here $n_a^i$ is the Killing vector corresponding to the specified rotation, $a=1,2,3$, and we used Gauss law to obtain the right-hand side.

We can now discuss the consequences of the vortex EFT for the CFT spectrum. To this aim, notice that the self-energy contribution $\sim\log Q$ in eq. \eqref{eqDelta3d} is proportional to $\sum_p q_p^2$ and implies that vortices with $|q|>1$ are energetically unfavored\footnote{Notice that vorticity is quantized $q_p\in\mathds{Z}$.}. The two main results of \cite{Cuomo} are:
\begin{itemize}
\item The lowest energy state for $\sqrt{Q}\ll J\leq Q$ consists of a vortex-antivortex pair rotating on the sphere, at a distance proportional to the spin $\Delta X/2=J/Q$ (see fig. \ref{fig0}). The scaling dimension of the corresponding operator reads
\begin{equation}\label{eqDelta3D2Vortices}
\Delta=\Delta_0(Q)+\frac{\sqrt{Q}}{3\alpha}\log\frac{J}{\sqrt{Q}}+2\tilde{\gamma}\sqrt{Q}+\mO\left(\sqrt{Q}\times\frac{Q}{J^2}\right).
\end{equation}
The leading correction to the ground state energy arises from the second term as a consequence of the logarithmic divergence of the vortex self-energy. This depends on the same coefficient $\alpha$ appearing in \eqref{eqDelta3d}.
The vortex mass contribution, given by the last term in \eqref{eqDelta3D2Vortices}, depends on a new coefficient and scales as the first subleading term in the ground state energy \eqref{eqDeltaQdDimensions}. Corrections to this formula arise from the particle velocities and the phonon field.
As $J\rightarrow \sqrt{Q}$, the vortices become relativistic and the derivative expansion breaks down. 
\begin{figure}
\centering
\includegraphics[scale=0.148,trim= 0 0 0 8cm]{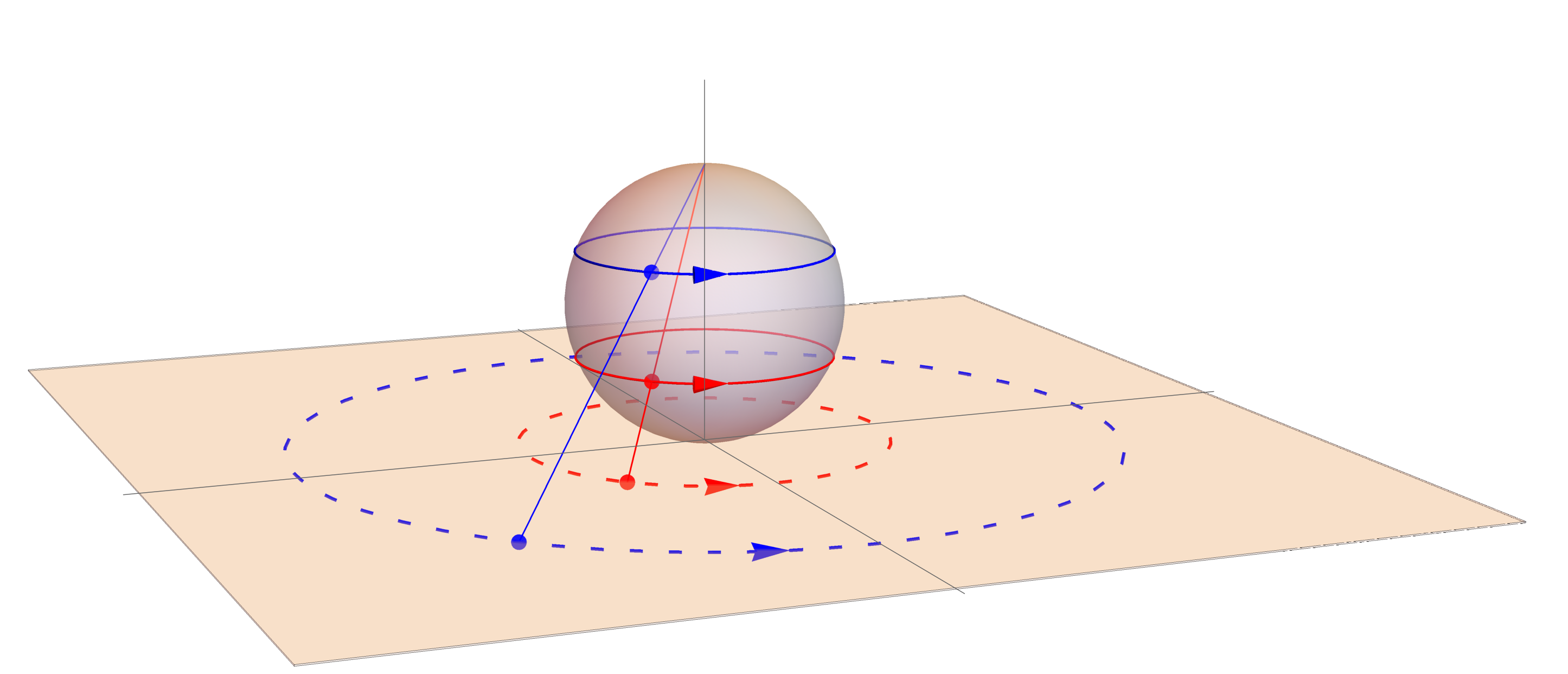}
\caption{A vortex-antivortex pair moving on the sphere at fixed distance; in the stereographic projection the motion corresponds to two circular orbits.}
\label{fig0}
\end{figure}
\item For $Q\ll J\ll Q^{3/2}$ the lowest energy state corresponds to a vortex crystal phase. Its energy is found approximating the vortex distribution as a continuous charge distribution $\rho(x)$ and then minimizing the energy at fixed angular momentum. The leading contribution to the energy arises from the electric field $|\vec{E}|\sim e^2|\rho|$ and reads 
\begin{equation}\label{eqDelta3DManyVortices}
\Delta=\Delta_0(Q)+\frac{1}{2\alpha}\frac{J^2}{Q^{3/2}}+
\mO\left(\frac{J^2}{Q^{3/2}}\times\frac{Q}{J},\frac{J^2}{Q^{3/2}}\times\frac{J^2}{Q^3}\right).
\end{equation}
corresponding to the charge density $\rho=\frac{3}{2\pi R^2}\frac{J}{Q}\cos\theta$. 
The second term in \eqref{eqDelta3DManyVortices} is the electrostatic energy of the crystal. The leading corrections arise from the vortex masses and the magnetic field fluctuations. The description holds as long as the electric field is subleading to the homogeneous monopole field $B$ and as long as the particle velocities are negligible. Using $|\vec{E}|\sim e^2|\rho|\sim J/\sqrt{Q}$ and $|\dot{\vec{X}}|\sim \left|\vec{E}\right|/B$, this sets the condition $J\ll Q^{3/2}$. \footnote{This is in agreement with the experimental fact that vortex crystals exist when the filling fraction $\nu= j_0/n_v$ is much bigger then one, where $n_v\sim |\rho|$ is the vortex density \cite{Moroz1,Moroz2}.}
\end{itemize}
For $J\ll \sqrt{Q}$ spinning operators are described by phonons \eqref{eqPhonon}.

\paragraph{Spin and additional degrees of freedom in the vortex cores}
In our analysis we implicitly assumed the simplest possible structure for the vortex cores, which do not carry any additional degrees of freedom on top. However, it is possible to imagine that the heavy particles have non zero spin, for instance. One might then wonder to what extent such an additional structure on the worldline can modify the results discussed so far. As this point was not addressed explicitly in \cite{Cuomo}, we would like to make few remarks in what follows.  

Consider for concreteness fermionic vortices with half-integer spin \footnote{We also assume parity; in this case, a Dirac particle in $2+1$ dimensions may have both positive or negative spin.} and focus on a state with a single vortex-antivortex pair. Spin degrees of freedom can be studied by adapting the formalism of \cite{spin1,spin2,spin3} to the conformal case. We report here only the key points, which can be easily understood via the analogy with the familiar case of a non-relativistic particle with spin in a magnetic field. Some details are given in appendix \ref{spinV}. 
Call $s^a_p$ the spin of the particle $p$ in embedding coordinates. The contribution of the latter to the angular momentum \eqref{eqAngularMom3d} is of the same order of other $\mO(1)$ contributions, proportional to the particle velocities, which we neglected; the expression \eqref{eqAngularMom3d} is thus not modified to the order of interest. Based on dimensional analysis and rotational invariance, we expect the existence of the spin $s^a_p$ to modify the energy \eqref{eqDelta3d} at leading order via a contribution of the kind
\begin{equation}\label{eqSpin3D}
\delta\Delta/R=\sum_p \frac{g_p}{\sqrt{2}}\sqrt{B}X^a_ps^a_p,\,
\end{equation}
where the coefficients $g_p$ can be interpreted as the magnetic moments of the vortices. This term indeed is just the Pauli interaction $\vec{B}\cdot\vec{s}/m$ between the spin and the magnetic monopole field for a particle of mass $m\sim\sqrt{B}$  \cite{spin0}. At large angular momentum, we can treat the positions of the vortices semiclassically as done before. Then, in the minimal energy state the spin vectors point in the direction which minimizes \eqref{eqSpin3D} at fixed positions for the vortices; this implies in particular that the doubling of the worldline degrees of freedom due to the spin does not lead to any degeneracy in the spectrum. Consequently, the result in eq. \eqref{eqDelta3D2Vortices} for the minimal energy state with angular momentum $\sqrt{Q}\ll J\leq Q$ is modified by the term
\begin{equation}
\delta\Delta=-\frac{g}{2}\frac{J}{\sqrt{Q}}\,,
\end{equation}
where we assumed all magnetic moments to be equal to $g>0$. Notice that this contribution is always subleading with respect to the second logarithmically enhanced term of eq. \eqref{eqDelta3D2Vortices} and it is at most of the same order of the vortex masses.

The conclusion of the previous analysis is general. The leading contribution to the angular momentum, eq. \eqref{eqAngularMom3d}, is unaffected by the presence of additional degrees of freedom characterizing the vortex cores. Similarly, the dominant contribution to the energy from the vortices always arises from the electrostatic interaction. Additional degrees of freedom might store energy in the vortex core, similarly to the vortex masses, and lead to the existence of new subleading corrections.

\section{Formulation of the EFT in four dimensions}\label{secEFTformulation}

\subsection{Dual gauge field}\label{secDualGaugeField}

As in $2+1$ dimensions, to write a local coupling between vortices and the superfluid we consider a dual description in terms of a gauge field.
Following the steps in sec. \ref{secVortices3D}, we rewrite the leading order Lagrangian \eqref{eqActionSuperfluid} in $d=3$ using a two form Lagrange multiplier $A_{\mu \nu}=-A_{\nu\mu}$:
\begin{equation}\label{eqDualization}
\mathcal{L}= c v^4-\frac{1}{4\pi}A_{\mu \nu} 
\frac{\epsilon^{\mu\nu\rho\sigma}}{\sqrt{g}}
\pd_\rho v_{\sigma},
\end{equation}
Integrating out $v_\mu$ then gives
\begin{equation}\label{eqActionPureGauge}
\mathcal{L}=-\kappa H^{4/3},\qquad
H_{\mu\nu\rho}=\pd_\mu A_{\nu\rho}
+\pd_\nu A_{\rho\mu}
+\pd_\rho A_{\mu\nu},
\end{equation}
where $H=\sqrt{-H_{\mu\nu\rho}H^{\mu\nu\rho}}$ and 
$\kappa=\frac{1}{16\pi^{4/3}}\left(\frac{3}{4c}\right)^{1/3}$. The $U(1)$ current provides the relation between $\chi$ and $A_{\mu \nu}$:
\begin{equation}\label{eqCurrent}
j^\mu=4 c (\pd\chi)^2\pd^\mu\chi=
\frac{1}{12\pi}\frac{\epsilon^{\mu\nu\rho\sigma}}{\sqrt{g}}H_{\nu\rho\sigma}.
\end{equation}
Consequently, the homogeneous charge density $\langle j^0\rangle=\frac{Q}{2\pi^2 R^3}$ in the vacuum translates into a constant background field:
\begin{equation}\label{eqMagneticBackground}
\langle H_{\eta \xi \phi}\rangle= -B \sin\eta\cos\eta,\qquad
B\equiv \frac{Q}{\pi R^3}.
\end{equation}
The cutoff of the theory \eqref{eqDualCutoff} is thus set by $B^{1/3}$ in the dual description.
The action \eqref{eqActionPureGauge} is often called of Kalb-Ramond type and is invariant under the gauge transformations
$A_{\mu\nu}\rightarrow A_{\mu\nu}+\pd_\mu \xi_\nu-\pd_\nu \xi_\mu$,
for an arbitrary vector $\xi_\mu$. The gauge redundancy allows imposing three gauge fixing conditions, since a gauge transformation generated by a total derivative $\xi_\mu=\pd_\mu\alpha$ acts trivially.

In the following, we shall be interested in fluctuations of the background \eqref{eqMagneticBackground}. It is thus convenient to expand the gauge field in a background value $\bar{A}_{\mu\nu}$ plus fluctuations:
\begin{equation}
A_{\mu\nu}=\bar{A}_{\mu\nu}+\delta A_{\mu\nu},
\end{equation}
where a possible choice is
\begin{equation}\label{eqGaugeChoice}
\bar{A}_{\eta\xi}=\bar{A}_{\eta\phi}=0,\qquad
\bar{A}_{\xi\phi}=-\frac{B}{2}\left(1-\cos^2\eta\right).
\end{equation}
Fluctuations are conveniently parametrized in terms of two three vectors $b^i$ and $a_i$ defined as:
\begin{equation}
\delta A_{ij}=\sqrt{g}\,\epsilon_{ijk}b^k,\qquad
\delta A_{0i}=a_i.
\end{equation} 
We partially 
fix the gauge requiring $\nabla_i A^{ik}=0$, which sets the curl of $b^i$ to zero. Then the Lagrangian to
quadratic order in the fluctuation reads:
\begin{equation}\label{eqPureGaugeActionFluctuations}
\mathcal{L}\simeq \frac{1}{4 e^2}f^2+\frac{1}{2 e^2}\left[
\dot{b}^i\dot{b}_i-\frac{1}{3}\left(\nabla_i b^i\right)^2\right],
\end{equation}
where $e^2=\frac{(\sqrt{6} B)^{2/3}}{8\kappa}$ and $f^2=f_{ij}f^{ij}$ with
\begin{equation}
f_{ij}=\pd_i a_j -\pd_j a_i.
\end{equation}
Following the gauge fixing, the field $b^i$ is purely longitudinal and corresponds to the phonon. Instead $a_i$ is a non-propagating degree of freedom, called the \emph{hydrophoton} since the residual $U(1)$ gauge invariance acts as $a_i\rightarrow a_i-\pd_i\xi_0$. Analogously to the Coulomb field in \eqref{eqDualGauge3d}, the hydrophoton does not correspond to a local field in the original description and provides the leading coupling to the vortices. 

\subsection{String-vortex duality}\label{secStringVortex}

Vortices in the dual description correspond to topological line defects, which are described as $1+1$ dimensional strings embedded in the $3+1$ dimensional spacetime \cite{Lund,Davis,Vortices}. The line element of a vortex $p$ is parametrized by $X_p^\mu(\tau,\sigma)$, where $\tau$ and $\sigma$ are the world-sheet coordinates. We use the words ``vortex'' and ``string'' interchangeably.
We also assume that no light degrees of freedom, besides the string coordinates, live on the worldsheet. 

The Lagrangian is required to be Weyl invariant and reparametrization invariant for both $\tau$ and $\sigma$ and is analogous to \eqref{eqAction3dVortices}. The lowest order terms are given by
\begin{multline}\label{eqAction1}
S=-\kappa\int d^4x \sqrt{g}H^{4/3}-\sum_p\lambda_p 
\int d\tau d\sigma A_{\mu\nu}\partial_\tau X_p^\mu\pd_\sigma X_p^\nu \\
-\sum_p\int d\tau d\sigma H^{2/3}\sqrt{|\text{det}(G_{\alpha\beta})|}
F_p\left[h_{\alpha\beta}G^{\alpha\beta}\right]
+\ldots\,.
\end{multline}
The first term was discussed in the previous section. The second term is the leading coupling between a string of vorticity $\lambda_p\in\mathds{Z}$ and the gauge field. The last term is the generalized Nambu-Goto (NG) action for the vortex; in appendix \ref{AppendixCoset} we derive its form via the coset construction. Here, the world-sheet metric is provided by:
\begin{equation}\label{eqStringMetricG}
G_{\alpha\beta}=g_{\mu\nu}\pd_\alpha X^\mu_p\pd_\beta X^\nu_p,\qquad\qquad
\alpha,\beta=\tau,\sigma.
\end{equation} 
Since the superfluid velocity breaks Lorentz invariance, one can construct another independent symmetric world-sheet tensor, which can be chosen as
\begin{equation}\label{eqStringMetrich}
h_{\alpha\beta}=\pd_\alpha X^\mu\pd_\beta X^\nu\,\frac{j_\mu j_\nu}{j^2}.
\end{equation}
In general the NG action contains an arbitrary function of $G^{\alpha\beta} h_{\alpha\beta}$, where $G^{\alpha\beta}$ is the inverse of $G_{\alpha\beta}$. Weyl invariance further fixes the power of $H$ which multiplies it. Finally dots in \eqref{eqAction1} stands for higher derivative terms.

Consider now the physical gauge $X_p^0=\tau$ for vortices. 
Using \eqref{eqGaugeChoice}, the second term in \eqref{eqAction1} is linear in time derivatives of the vortex line. As we will self-consistently see in the next section, this implies that vortices move with drift velocity $|\dot{\vec{X}}|\sim f/B\sim  B^{-1/3}$. Then, similarly to what we argued below \eqref{eqAction3dVortices},
terms of the kind $\dot{\vec{X}}\cdot\dot{\vec{X}}$ in the NG action can be treated as higher derivatives and we neglect them. The coupling of the phonon field to the strings is also negligible to leading order. In this regime, the action reduces to
\begin{multline}\label{eqActionFluctuation}
S\simeq \frac{1}{e^2}\int d^4x \left\{\frac14 f^2
+\frac{1}{2}\left[
\dot{b}^i\dot{b}_i-\frac{1}{3}\left(\nabla_i b^i\right)^2\right]\right\}
\\
-\sum_p 
\int d\tau d\sigma \left[\lambda_p\left(\bar{A}_{i j}\partial_\tau X^i\pd_\sigma X^j + a_{i}\pd_\sigma X_p^i \right)
+\gamma_p B^{2/3} (\pd_\sigma X)\,\right],
\end{multline}
where $\gamma_p=6^{1/3}F_p\left(1\right)$ and we define
\begin{equation}
(\pd_\sigma X)=\sqrt{|g_{ij}|\pd_\sigma X^i_p\pd_\sigma X^j_p}\,.
\end{equation}
Notice that the phonon spectrum \eqref{eqPhonon} to leading order is not affected by the presence of vortices.

\section{Results of the EFT}\label{secVortexEFT}

\subsection{Classical analysis}

From the leading order action \eqref{eqActionFluctuation} the following equations of motion for the hydrophoton and the strings are derived
\begin{equation}\label{eqEOMfij}
-\frac{1}{e^2}\nabla_i f^{ij}=\sum_p \mJ_p^{j}\equiv  \sum_p
\lambda_p
\int d\sigma\,
\pd_\sigma X_p^j
\frac{\delta^3(x^i-X^i_p)}{\sqrt{g}},
\end{equation}
\begin{equation}\label{eqEOMstring}
\lambda_p\left(f_{ik}-B\sqrt{g}\epsilon_{ijk}\dot{X}^j_p
\right)\pd_\sigma X^k_p
=\gamma_p B^{2/3}|g_{ij}|\frac{D}{D\sigma}\left[
\frac{\pd_{\sigma} X^j}{(\pd_\sigma X)}\right].
\end{equation}
Eq. \eqref{eqEOMfij} is analogous to Amp\`{e}re's circuital law in magnetostatic, a vortex acting as an electric current $\mJ^i_p$ sourcing the field $f^{ij}$. 
Eq. \eqref{eqEOMstring} is the string equation of motion. Notice that it is first order in time derivatives and implies that vortices move with drift velocity
$|\dot{\vec{X}}|\sim f/B\sim B^{-1/3}$. The right-hand side arises from the NG action and it is proportional to the covariant derivative of the line element $\frac{D}{D\sigma}\left[
\frac{\pd_{\sigma} X^j}{(\pd_\sigma X)}\right]$; the left-hand side comes from the minimal coupling to the gauge field. 

As in sec. \ref{secVortices3D} the electrostatic problem required the net charge on the sphere to be zero, the $3+1$ dimensional magnetostatic problem defined by \eqref{eqEOMfij} and \eqref{eqEOMstring} requires zero vorticity flux on every closed surface. To this aim, we only consider closed strings. 
This point is perhaps more easily understood considering a vortex configuration in the scalar description \eqref{eqActionSuperfluid} \cite{Komargodski:2018odf}. In that language, this is a configuration where the value of the field changes from $\pi=0$ to $\pi=2 \pi\lambda$ from below to above of a certain $2d$ space surface, where $\lambda\in\mathds{Z}$ ($\lambda\neq 0$) is the vorticity. The vortex is just the boundary of this surface. As $S^3$ is a compact manifold, the string must form a closed curve. In this picture it is also clear that a closed vortex configuration cannot break into an open string. \footnote{More formally, the surface described above is the object charged under the 2-form symmetry associated with the current $J_{\mu\nu\rho}=\sqrt{g}\epsilon_{\mu\nu\rho\sigma}\pd^\sigma\chi\propto H_{\mu\nu\rho}/H^{2/3}$ \cite{Gaiotto:2014kfa}; conservation of this current, associated with the winding number of the Goldstone, forbids the breaking of a closed string. } A similar reasoning can be used in $2+1$ dimensions to argue that the net \emph{electric} charge on the sphere must vanish.

The energy and angular momentum associated to solutions of the EOMs are computed from the stress energy tensor $T_{\mu\nu}=\frac{2}{\sqrt{g}}\frac{\delta S}{\delta g^{\mu\nu}}$:
\begin{multline}
T_{\mu\nu}=\frac{\kappa}{H^{2/3}}\left(4 H_{\mu\sigma\rho} H_{\nu}^{\,\,\sigma\rho}+
 g_{\mu\nu} H^{2}\right)\\
+\sum_p \gamma_p B^{2/3} \int d\tau d\sigma\frac{\delta^4(x^\mu-X^\mu_p)}{\sqrt{g}}
 \sqrt{|\text{det}(G_{\alpha\beta})|}\,G^{\alpha\beta}
\,\pd_\alpha X^\sigma_p\pd_\beta X^\rho_p \,g_{\sigma\mu}g_{\rho \nu}.
\end{multline}
The classical energy of the state is found from 
\begin{equation}\label{eqEnergy}
E=\frac{\Delta}{R}=\frac{3 Q^{4/3}}{8 \pi ^{2/3} c^{1/3}R}+
\frac{1}{4e^2}\int d^3 x\sqrt{g}f^2+\sum_p
\gamma_p B^{2/3}\int d\sigma(\pd_\sigma X).
\end{equation}
The first term is the energy of the homogenous ground state. The second term is the energy stored in the \emph{magnetostatic} field $f_{ij}$ created by the vortices. Finally, the last term is the energy contribution from the
tension and
is proportional to the length of the string $L_p$. 

Eq. \eqref{eqEOMfij} gives the field $a_i$ in terms of the string current:
\begin{equation}
a_i(x)=e^2\sum_p\int d^3x' G_{ij}(x,x')\mJ_p^j(x'),
\end{equation}
where $G_{ij}(x,x')$ is the photon propagator on $S^3$. In appendix \ref{AppProp} it is shown that the photon Green function on $S^d$ takes the form
\begin{equation}\label{eqPhotonProp}
G_{ij'}(x,x')=-\left(\pd_i\pd_{j'}u(x,x')\right)F(u(x,x')),\qquad u=\frac{1}{2}\Delta X^2(x,x')
\end{equation}
where $\Delta X^2$ is the chordal distance between two points in embedding space, and
\begin{equation}
F(u)=\frac{\Gamma(d-2)}{(4\pi)^{\frac{d}{2}}\Gamma\left(\frac{d}{2}\right)
R^{d-2}}
\,_2 F_1\left(1,d-2;\frac{d}{2};1-\frac{u}{2 R^2}\right).
\end{equation}
Then the scaling dimension of the corresponding operator can be written as
\begin{equation}\label{eqDimension}
\Delta=\alpha Q^{4/3}+\frac{Re^2}{2}\sum_{p,p'}
\int d^3x\sqrt{g} \int d^3x'\sqrt{g'}\mJ^{j}_p(x)G_{jk'}(x,x')\mJ^{k'}_{p'}(x')
+\sum_p
\gamma_p R\,B^{2/3}L_p,
\end{equation}
where $\alpha=\frac{3 }{8 \pi ^{2/3} c^{1/3}}$. Notice the analogy with the structure of \eqref{eqDelta3d}. 

The angular momentum (in units of $1/R$) of the corresponding state can be computed similarly:
\begin{equation}\label{eqAngularMom}
J_{ab}=\frac{RB}{2 e^2} \int d^3 x\sqrt{g} \,\mathit{n}^i_{ab}\epsilon_{i jk}\sqrt{g}f^{jk},
\end{equation}
where $\mathit{n}_{ab}$ is the Killing vector corresponding to a rotation in the $(X_a,X_b)$ plane. Using Amp\`{e}re's law \eqref{eqEOMfij} and Stoke's theorem, it is conveniently rewritten as
\begin{equation}\label{eqAngularMomArea}
\frac12 J_{ab}\epsilon_{abcd}=
-\frac{RB}{2}\sum_p\lambda_p \int d\sigma_p \left[X_c^p(\pd_\sigma X_d^p)-
X_d^p(\pd_\sigma X_c^p)\right]
=-RB\sum_p\lambda_p\int dX^p_c\wedge dX^p_d,
\end{equation}
where $X^p_a$ are the vortex coordinates in the $\mathds{R}^4$ embedding of $S^3$. The last equation on the right-hand side is a formal notation for the area enclosed by the vortex projection in the $(X_c,X_d)$ plane.

In the following we will study simple specific configurations.

\subsection{Vortex rings}\label{secDrift}

In nature, vortices often have a ring shape and move with a constant speed inversely proportional to the radius \cite{Vortex0}. It is hence natural to look for vortex ring solutions of the EOMs \eqref{eqEOMfij} and \eqref{eqEOMstring}. As we will see, a vortex ring generalizes the vortex-antivortex configuration in fig. \ref{fig0}. 

The simplest configuration one can study is a slowly moving vortex ring with unit negative charge $\lambda=-1$.
We pick the gauge $\xi=\sigma$ and consider a radius $r R\leq R$ ring in the $(X_1,X_2)$ plane in embedding space. The EOMs implies that the ring rotates with constant drift velocity $v$ in the $(X_3,X_4)$ plane:
\begin{equation}\label{eqSingleVortexConfig}
X_1^2(t,\sigma)+X_2^2(t,\sigma)=R^2\sin^2\eta(t,\sigma)=R^2r^2=const.\qquad
\phi(t,\sigma)= vt+const.\;.
\end{equation}
The precise value of $v$ is fixed by eq. \eqref{eqEOMfij}. From eq. \eqref{eqAngularMomArea} it follows that the only nonvanishing component of the angular momentum is given by:
\begin{equation}
J_{34}=Q r^2 .
\end{equation}
\begin{figure}
\centering
\includegraphics[scale=0.17,trim= 10cm 5cm 0 10cm]{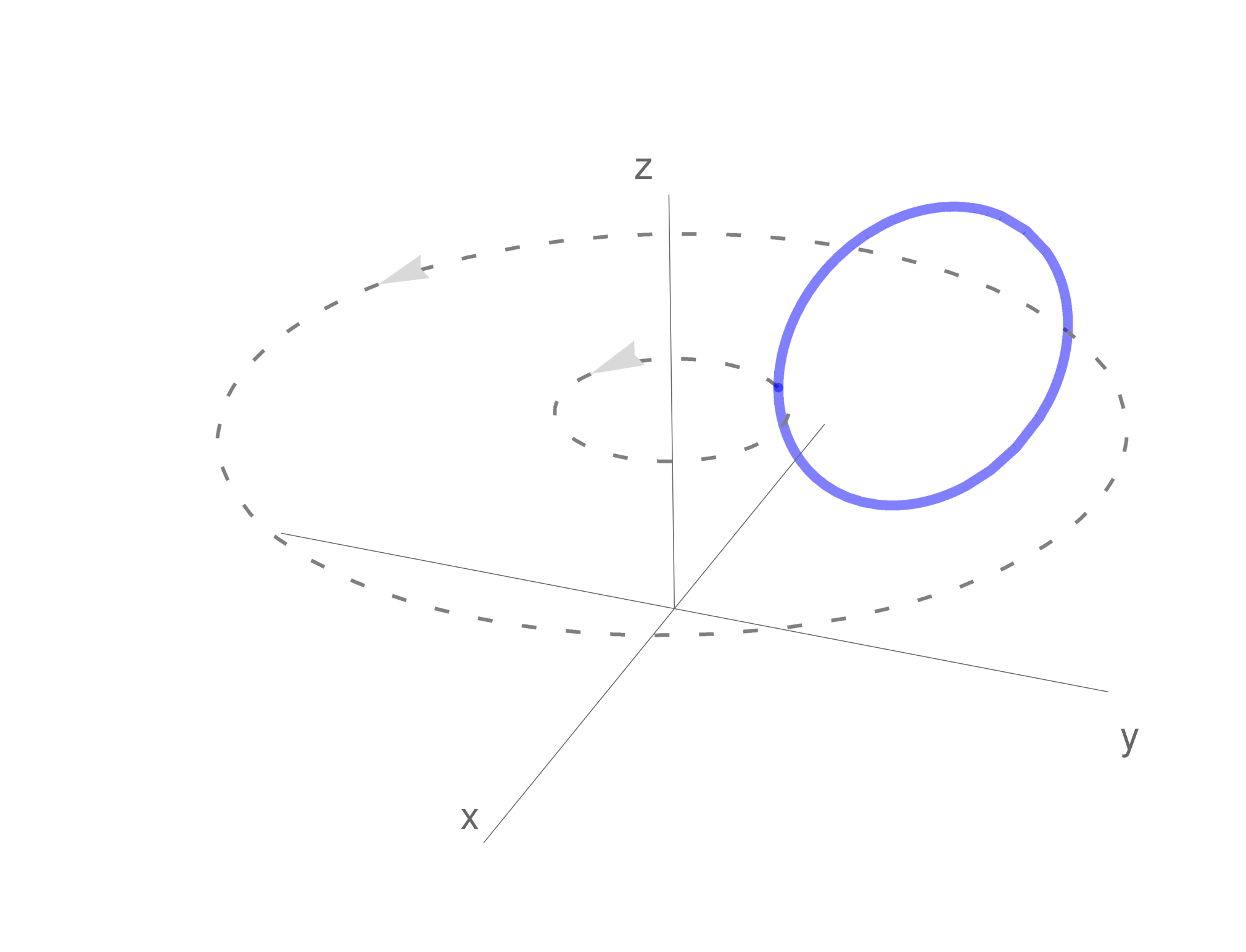}
\caption{The vortex ring orbit in stereographic coordinates.}
\label{fig1}
\end{figure}In figure \ref{fig1} the motion is depicted in stereographic coordinates, defined by the relation $(x,y,z)=\frac{1}{1+X_1}\left(X_3,X_4,X_2\right)$. Eq. \eqref{eqSingleVortexConfig} corresponds to a ring orbiting around the $z$ axis; as the angular momentum is increased, the ring size increases and its velocity decreases. For $r\rightarrow 1$ the surface embedded by the ring in the stereographic projection extends to cover the whole plane and the vortex lies statically on the geodesic corresponding to the $z$ axis. Fig. \ref{fig1} qualitatively generalizes the $2+1$ dimensional motion depicted in fig. \ref{fig0}.  

Using \eqref{eqDimension} we can calculate the energy of this configuration as:
\begin{equation}\label{eqEnergy1VortexPrePre}
E=\alpha Q^{4/3}/R+\frac{e^2R}{2}\iint d\xi d\xi'\mJ^{i}(\xi)  G_{ij}\left(x(\xi),x(\xi')\right) \mJ^{j}(\xi')+\gamma  B^{2/3}2\pi rR,
\end{equation}
The only nontrivial contribution arises from the second term, corresponding to the magnetostatic self-energy of the string. It diverges due to the short distance behaviour of the hydrophoton propagator. We regulate the calculation working in $d+1$ spacetime dimensions, as explained in appendix \ref{AppMagnetostaticEnergy}; the result is
\begin{multline}\label{eqEnergy1VortexPre}
E=\alpha Q^{4/3}/R+ e^2\pi R \Big\{
\frac{r}{2 \pi  (3-d)}+
\frac{r \left[\log \left(4 \pi  r^2 B^{2/3}R^2\right)+\frac92+\frac{1}{3}\log 6-\gamma_E -
2 \psi \left(\frac32\right)\right]}{4 \pi }
\\
-\frac{r}{2\pi}\log (r+1)-\frac{1}{\pi}\log (r+1)\Big\}+
\gamma  B^{2/3}2\pi rR.
\end{multline}
Details of the computation are given in appendix \ref{App_self_vortex}.
There is a divergent piece for $d\rightarrow 3$ proportional to the vortex length, which renormalizes the string tension. The contribution logarithmically enhanced by the cutoff $\sim  e^2 r\log \left(r^2 B^{2/3}\right)$ can be seen as a consequence of the renormalization group running of $\gamma$ induced by the hydrophoton \cite{Vortices}. Collecting everything,
the scaling dimension \eqref{eqDimension} for a vortex ring state reads
\begin{equation}\label{eqDelta1Vortex}
\Delta=\alpha Q^{4/3}+\Delta_{V}(Q,J_{34}),
\end{equation}
where we isolated the vortex contribution to the energy:
\begin{multline}\label{eqDeltaV}
\Delta_{V}(Q,J)=\frac{3}{8\alpha}Q^{1/6}J^{1/2}\log \left(  J/Q^{1/3}\right)
-\frac{3}{4\alpha}Q^{1/6}J^{1/2}\log \left(1+\sqrt{J/Q}\right)\\
-\frac{3}{2\alpha}Q^{2/3}\log \left(1+\sqrt{J/Q}\right)
+\tilde{\gamma}Q^{1/6}J^{1/2}.
\end{multline}
Here $\tilde{\gamma}$ is a finite new coupling which absorbs all contributions proportional to $r$ in \eqref{eqEnergy1VortexPre}.  As in \eqref{eqDelta3D2Vortices}, the leading contribution arises because of the classical running of the tension induced by the magnetostatic self-energy and is given by the first term in \eqref{eqDeltaV}.
For $J\ll Q$, the other contributions can be expanded in powers of the vortex length and to leading order effectively scale as $Q^{1/6}J^{1/2}$. Physically, this is understood noticing that the vortex energy \emph{density} is set by $e^2\sim Q^{2/3}$, hence for short vortices the energy can be estimated as the length times the energy density (neglecting the logarithmic running of the tension): $2\pi rR\times e^{2/3}\sim  Q^{1/6}J^{1/2}$. However, as $J\rightarrow Q$ the functional dependence of the second and third term in eq. \eqref{eqDeltaV} deviates from this expectation, as a consequence of the vortex finite size. 

As the ring radius is decreased to inverse cutoff length 
$r\rightarrow 1/(\Lambda R)$, corresponding to $J_{34}\rightarrow  Q^{1/3}$, the magnetostatic field $f\sim e^2/(Rr)$ becomes of the same order of the background field $B$ and the vortex velocity approaches the relativistic regime. Hence subleading contributions to \eqref{eqActionFluctuation} become unsuppressed and the EFT breaks down. 

Eq. \eqref{eqDelta1Vortex} can be identified as the minimal energy state at fixed angular momentum in its regime of validity.

We now study states with two vortices, one laying on the $(X_1,X_2)$ plane and the other on the $(X_3,X_4)$ plane in embedding space. Because of \eqref{eqAngularMomArea}, these configurations are associated to operators in mixed symmetric representations of the $SO(4)$ group. 

Consider first a radius $R$ ring in the $(X_1,X_2)$ plane interacting with a ring of arbitrary size in the $(X_3,X_4)$ plane. In this geometry, the interaction does not affect the equations of motion and the solution takes a simple form
\begin{equation}
\begin{array}{ccc}
\text{vortex }1: & X_1^2(t,\sigma_1)+X_2^2(t,\sigma_1)=R^2, &
\sigma_1=\xi_1  ;   \\
\text{vortex }2: &
\cos^2\eta_2(t,\sigma_2)=r_2^2,\qquad
\xi_2(t,\sigma_2)= v_2t,\qquad
&\sigma_2=\phi_2    .
\end{array}
\end{equation}
Focussing on negative unit charge vortices $\lambda_1=\lambda_2=-1$, this configuration corresponds to an operator in a mixed symmetric representation with spin given by
\begin{equation}
J_{34}=  Q ,\qquad
J_{12}=Q r_2^2.
\end{equation}
Since the \emph{electric} currents $\mJ^i$ sourced by the strings are orthogonal, the corresponding scaling dimension is found analogously to \eqref{eqDelta1Vortex}:
\begin{equation}\label{eqDelta2Vortices}
\Delta=\alpha Q^{4/3}+\Delta_{V}(Q,J_{34})+\Delta_{V}(Q,J_{12}).
\end{equation}
To leading order, a similar solution exists for $0\leq (1-r_1^2)\ll (R\Lambda)^{-2}$, hence for $0\leq Q- J_{34}\ll Q^{1/3}$. As before, the consistency of the EFT requires $J_{12}\gg Q^{1/3}$.

In general, the mutual interaction affects non trivially the motion of the two vortex rings. One can, however, identify the logarithmically enhanced contributions analogous to the first term in \eqref{eqDeltaV} just from the free action. These indeed arise from the running of the tension induced by the hydrophoton contribution to the vortex self-energy. For $Q^{1/3}\ll J_{12},J_{34}\leq Q$, the leading contribution to the energy reads:
\begin{equation}\label{eqDelta2VorticesGuess}
\Delta= \alpha Q^{4/3}+
\frac{3}{8\alpha}Q^{1/6}\left[J_{34}^{1/2}\log \left(  J_{34}/Q^{1/3}\right)
+J_{12}^{1/2}\log \left(  J_{12}/Q^{1/3}\right)\right].
\end{equation}
This result holds as long as the minimal distance $d$ between the two vortices is larger than the inverse of the cutoff:
\begin{equation}\label{eqVortexDistance}
\frac{d^2}{R^2}\sim\frac{\left(J_{12}+J_{34}-Q\right)^2}{J_{12}J_{34}}\gg \frac{1}{Q^{2/3}}.
\end{equation}

\subsection{Vortex crystals}\label{secMultiVortices}

Since the magnetostatic self-energy of a single vortex is proportional to $\lambda^2$, strings with $|\lambda|\geq 1$ are energetically unfavored. Hence the minimal energy state for values of the angular momentum $J_{34}\gg Q$ is made by $n\gg 1$ vortices. We then approximate the vortex distribution with a continuous current density $\mJ^i(x)$. The corresponding state
is found minimizing the energy \eqref{eqEnergy} at fixed angular momentum \eqref{eqAngularMom}, giving the following density profile:
\begin{equation}\label{eqDensity1}
\mJ^{\xi}=\frac{2}{\pi R^2}\frac{J_{34}}{Q},\qquad \mJ^\phi=\mJ^\eta=0.
\end{equation} 
The leading contribution to the energy comes from the magnetostatic field
and reads
\begin{equation}\label{eqVortexCrystal1}
\Delta=\alpha Q^{4/3}+\frac{3}{4\alpha}\frac{J_{34}^2}{Q^{4/3}}.
\end{equation}
Physically, this state corresponds to a vortex crystal \cite{VortexCrystalNature,Moroz1,Moroz2}.
When $J_{34}\rightarrow Q^{4/3}$, the magnetic field $f$ approaches $B$, vortices become relativistic and the EFT breaks down. 

Similarly, the ground state for $Q\ll J_{34},J_{12}\ll Q^{4/3}$ is provided by a vortex crystal, whose current density and energy are given by
\begin{equation}\label{eqDensity2}
\mJ^{\xi}=\frac{2}{\pi R^2}\frac{J_{34}}{Q},\qquad 
\mJ^\phi=\frac{2}{\pi R^2}\frac{J_{12}}{Q},\qquad
\mJ^\eta=0,
\end{equation}
\begin{equation}\label{eqVortexCrystal2}
\Delta=\alpha Q^{4/3}+\frac{3}{4\alpha}\frac{J_{34}^2+J_{12}^2}{Q^{4/3}}.
\end{equation}

\subsection{Quantization and Kelvin waves}\label{secKelvinWaves}

Vortices in four dimensions are extended objects and can thus propagate \emph{Kelvin} waves on them \cite{Vortices}. The corresponding states are associated to primary operators in the CFT. To study them,
we consider a single string of vorticity $\lambda=-1$. It is convenient to parametrize its coordinates via the following variables:
\begin{align}
z(t,\sigma)&=X_1(t,\sigma)+iX_2(t,\sigma)=R\sin\eta(t,\sigma) e^{i\xi(t,\sigma)},
\\
w(t,\sigma)&=X_3(t,\sigma)+iX_4(t,\sigma)=R\cos\eta(t,\sigma) e^{i\phi(t,\sigma)}. 
\end{align}
These are related through the constraint $|z|^2+|w|^2=1$.
We pick the gauge $\xi=\sigma$ and $t=\tau$. Integrating out explicitly the hydrophoton from eq. \eqref{eqActionFluctuation}, we find the single vortex action as
\begin{multline}\label{eqSingleVortexActionGeneral}
S_{1-vortex}=\int dt d\sigma\left[i \frac{B}{2}
w^*\dot{w}-\gamma B^{2/3}
\sqrt{|\partial_{\sigma}z|^2+|\partial_{\sigma}w|^2}
\right]\\
+\frac{e^2}{4}\int dt d\sigma d\sigma'
\left(\pd_\sigma \pd_{\sigma'} \Delta X^2(\sigma,\sigma')\right)F\left(
\frac{\Delta X^2(\sigma,\sigma')}{2R^2}\right),
\end{multline}
where $F$ is given in \eqref{eqPhotonProp}.
Eq. \eqref{eqSingleVortexActionGeneral} can be seen as the (nonlocal) action of a complex field $w(t,\sigma)$ living on $\mathds{R}\times S^1$. 
It is manifestly invariant under the action of the unbroken rotation generators $J_{34}$, corresponding to rotations around the vortex $w\rightarrow e^{i\alpha} w$, and $J_{12}$, corresponding to translations along the string $\sigma\rightarrow\sigma+\alpha$.

We expand for small fluctuations around the background $w=0$, which describes a radius $R$ ring in the $(X_1,X_2)$ plane with $J_{34}=Q$. The action to quadratic order reads:
\begin{equation}\label{eqSingleVortexActionSmall}
S_{1-vortex}\simeq\int dt d\sigma\left[i\frac{B}{2}w^*\dot{w}-\gamma B^{2/3}-\frac{\gamma B^{2/3}}{2}
|\partial_{\sigma}w|^2+\frac{\gamma B^{2/3}}{2}|w|^2
\right]+S_{non-local}^{(2)},
\end{equation}
where $S_{non-local}^{(2)}$ is found expanding the second line in \eqref{eqSingleVortexActionGeneral}. It follows that the vortex is quantized as a standard non-relativistic field:
\begin{equation}
w(t,\sigma)=\sqrt{\frac{2}{ B}}\sum_{n=-\infty}^\infty\frac{a_n}{2\pi}e^{-i\omega_n t+in\sigma},\qquad
[a_n,a^\dagger_m]=2\pi\delta_{nm}.
\end{equation}
As usual the $a_n$ annihilate the vacuum $a_n\ket{0}=0$, and thus so does $w(t,\sigma)$.
The proper frequencies $\omega_n$ are computed in appendix \ref{App_LinearKelvon} and read
\begin{equation}\label{eqKelvonFreq}
R\omega_n\equiv \Delta_{k}(n)=\frac{\pi(n^2-1)}{Q^{1/3}}\left\{
\frac{3}{8\alpha}
\left[\log Q^{2/3}-2 \psi \left(\frac{n+1}{2}\right)-2\gamma_E-1-\log 64\right]+\tilde{\gamma}
\right\}.
\end{equation}
Notice that the $n=0$ mode decreases the energy, while the $n=\pm 1$ modes have $\omega_{\pm 1}=0$. This can be understood from the expression of the angular momentum in terms of ladder operators at order $\mO(Q^0)$.
The rotations generated by $J_{12}$ and $J_{34}$ are linearly realized and their generators are quadratic in terms of ladder operators:
\begin{equation}\label{eqJquantum1}
J_{34}=Q-\sum_n \frac{a^\dagger_n a_n}{2\pi},\qquad
J_{12}=\sum_n n\frac{a^\dagger_n a_n}{2\pi}.
\end{equation}
The string realizes \emph{nonlinearly} the full rotation group. As a consequence, the broken components of the angular momentum are linear in the $n=\pm 1$ annihilation and creation operators:
\begin{equation}\label{eqJquantum2}
\begin{gathered}
J_{23}+J_{14}=-\sqrt{\frac{Q}{2\pi}}\left(a_{-1}+a^\dagger_{-1}\right),
\qquad
J_{23}-J_{14}=\sqrt{\frac{Q}{2\pi}}\left(a_{1}+a^\dagger_{1}\right),\\
J_{31}+J_{24}=i\sqrt{\frac{Q}{2\pi}}\left(
a_{-1}-a^\dagger_{-1}\right),
\qquad
J_{31}-J_{24}=i\sqrt{\frac{Q}{2\pi}}
\left(a_{1}-a^\dagger_{1}\right).
\end{gathered}
\end{equation}
\begin{figure}
\centering
\includegraphics[scale=0.17, trim= 0 0 0 6cm]{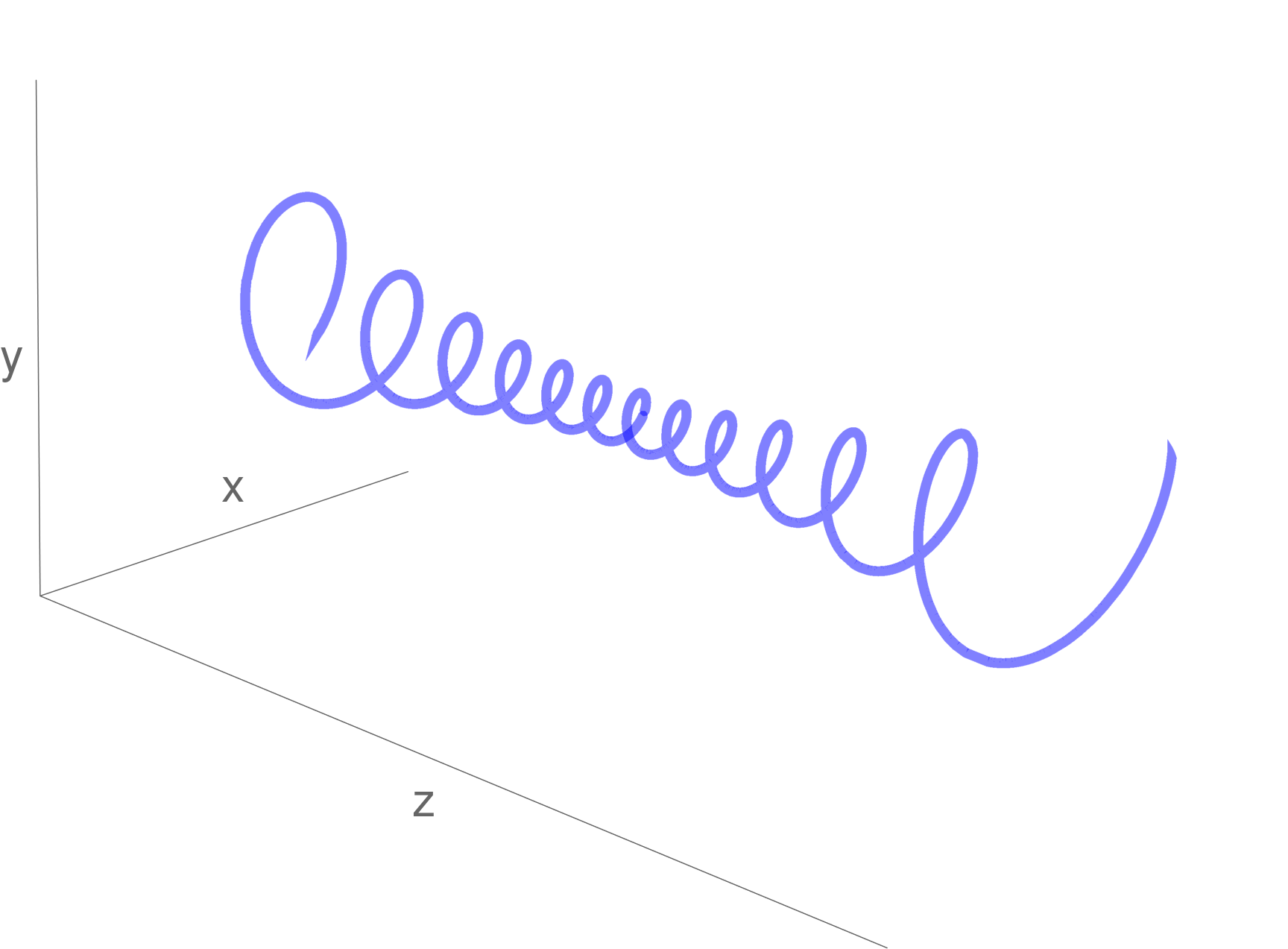}
\caption{A Kelvin wave in stereographic coordinates $(x,y,z)=\frac{1}{1+X_1}\left(X_3,X_4,X_2\right)$.}
\label{fig2}
\end{figure}From \eqref{eqJquantum1} we see that the $n=0$ mode decreases $J_{34}$ (and the radius of the vortex) by one unit, hence it corresponds to the quantization of the classical ring solution discussed in \ref{secDrift}.
Eq. \eqref{eqJquantum2} implies that the $n=\pm 1$ modes do not correspond to new states, but describe rotations of the string orientation and therefore have vanishing frequency. In this sense, their role is analogous to that of the $J=1$ phonons in \eqref{eqPhonon}, describing descendants of the ground state. 

The modes with $|n|\geq 2$ correspond to new solutions and are interpreted as Kelvin waves propagating on the vortex; in the CFT they correspond to operators with the following quantum numbers
\footnote{In $\Delta$ we neglect a $\sim Q^{-1/3}$ contribution from the vortex Casimir energy; this term does not depend on $J_{12}$ and can be thought as a subleading correction to $\Delta_V$.} 
\begin{equation}
J_{34}=Q-1,\qquad
J_{12}=n,\qquad
\Delta=\alpha Q^{4/3}+\Delta_V(Q,Q)+\Delta_k(n).
\end{equation}
As shown in fig. \ref{fig2}, a Kelvin wave in stereographic coordinates takes the form of a solenoid, trapping the magnetic field inside. The string undergoes a helical motion analogous to the one of a wine opener.

Notice that Kelvin waves carry less energy than phonons with the same angular momentum \eqref{eqPhonon}. It follows that a state obtained acting on the vacuum as
\begin{equation}
(a^\dagger_0)^{m}a^\dagger_n\ket{0}\equiv\ket{J_{34}=Q-m-1,J_{12}=n}
\end{equation}
is the minimal energy state for the specified value of the angular momentum.

This description applies in the linear regime $m+1=Q-J_{34}\ll Q$.
When $n=J_{12}\rightarrow Q^{1/3}$ higher derivative terms become unsuppressed and the EFT breaks.

\subsection{Higher order corrections}

Corrections arise from higher derivative terms we neglected in \eqref{eqActionFluctuation} and are suppressed by powers of the cutoff scale \eqref{eqDualCutoff}. Following \cite{Cuomo}, here we comment on their form. 

The first class of corrections was discussed in \cite{Hellerman,MoninCFT} and arises considering the effect of curvature terms in the superfluid and vortex action; these corrections are controlled by the sphere radius and hence scale as $1/(\Lambda R)^2\sim 1/Q^{2/3}$. They are also present in the absence of vortices and provide the subleading terms in \eqref{eqSummaryDelta0}. 

Focus now on the single vortex state described in sec. \ref{secDrift}. We find corrections controlled by the vortex length $L\sim\sqrt{J_{34}/Q}$, which hence scale as $1/(\Lambda L)^2\sim Q^{1/3}/J_{34}$ (we assume parity invariance, hence they depend only on $L^2$). They arise from the terms we neglected in the NG action to write \eqref{eqAction1} and are proportional to $(\nabla_i b^i)/B$, $f^2/B^2$ and $\dot{\vec{X}}^2$. Higher derivatives of the string line element as well as the phonon contribution to the energy \eqref{eqActionPureGauge} belong to the same class. Similarly, there are corrections of the form $Q^{1/3}/J_{34},\;Q^{1/3}/J_{12}$ to eq. \eqref{eqDelta2Vortices} for a two vortex state. Notice that the subleading $Q^{2/3}$ term in the ground state energy is bigger than the vortex contribution \eqref{eqDeltaV} for $Q^{1/3}\ll J_{34}\ll Q$. The latter gives instead the leading contribution for $J_{34}\sim Q$. The vortex contribution is anyway functionally distinguished from the ground state energy correction and is thus always calculable.

Let us now turn our attention to the Kelvin waves discussed in \ref{secKelvinWaves}. The same corrections discussed for a vortex ring exists in this case. Furthermore, for $n\gg1$ higher derivative corrections to the single vortex action \eqref{eqSingleVortexActionSmall} become important. As typical for a non-relativistic field, these arise due to terms with two time derivatives, or, equivalently, with four space derivatives (suppressed by an extra $H^{-2/3}$ factor by Weyl invariance) and scale as $n^2/Q^{2/3}=J_{12}^2/Q^{2/3}$. 
Notice that the relative corrections to the ground state energy of the vortex are bigger than the Kelvin wave energy \eqref{eqKelvonFreq} for $J_{12}^2/Q^{1/3}\lesssim Q^{1/2}/J_{34}^{1/2}$; however, these corrections are independent of $J_{12}$, which enters only through \eqref{eqKelvonFreq}.

Finally, the leading corrections to the energy of the vortex crystals states discussed in \ref{secMultiVortices} arise both from the phonon contribution to the energy, which is proportional to $(\nabla_i b^i)^2/f^2\sim\left(J_{ab}/Q^{4/3}\right)^2$, and from the free tension contribution, which gives $Q/J$ corrections using \eqref{eqDensity1} or \eqref{eqDensity2}. Here $J_{ab}$ stands for both $J_{34}$ and/or $J_{12}$ depending on the state. 

\section{Correlators}\label{secCorrelators}

We now turn our attention to the study of correlators. As in \cite{Cuomo}, the most natural correlation function\footnote{To leading order, scalar insertions read as in the homogeneous phase \cite{MoninCFT}.} which can be studied corresponds to a current insertion within two equal vortex states. In the EFT, this is determined through the following relations:
\begin{equation}\label{eqCorrelators}
\langle j_0\rangle= \frac{Q}{2\pi^2 R^3},\qquad
\langle j_{\phi}\rangle= \frac{\sqrt{g}}{2\pi}f^{\eta\xi},\qquad
\langle j_{\xi}\rangle= -\frac{\sqrt{g}}{2\pi}f^{\eta\phi}.
\end{equation}
The hydrophoton field is obtained from \eqref{eqEOMfij}, which,  in analogy with Amp\`{e}re's law, can be conveniently rewritten in integral form as
\begin{equation}
\frac12 \oint_{\mathcal{C}} dx^i \epsilon_{ijk}\sqrt{g} f^{jk}=-e^2 \lambda_{enc},
\end{equation}
where $\lambda_{enc}$ is the vorticity flux through the surface enclosed by the curve $\mathcal{C}$. Using this relation, eq.s \eqref{eqCorrelators} can be used to make nontrivial predictions about the OPE coefficients of the theory. 

Consider first the traceless symmetric state corresponding to a radius $R$ vortex in the $(X_1,X_2)$ plane, which has $J_{34}=Q$ and $J_{12}=0$. For this state, eq. \eqref{eqCorrelators} reads:
\begin{equation}\label{eqCorrelatorQQ}
\langle j_0\rangle= \frac{Q}{2\pi^2 R^3},\qquad
\langle j_{\phi}\rangle= \frac{e^2}{4\pi^2 R},\qquad
\langle j_{\xi}\rangle= 0.
\end{equation}
The expectation value of a spin-1 parity even conserved operator in a traceless symmetric state $\ket{(J,J),J_{34}=2J,J_{12}=0}$ is \cite{SpinningConformalCorrelators}:
\begin{equation}\label{eqCorrelatorTTCFT}
\begin{gathered}
\braket{(J,J),2J,0|j_{0}(\eta,\xi,\phi)|(J,J),2J,0}
=R^{-3}\sum_{m=0}^{2J}a_m\sin^{2m}\eta,
\\ 
\braket{(J,J),2J,0|j_{\phi}(\eta,\xi,\phi)|(J,J),2J,0}
=R^{-3}\sum_{m=0}^{2J}b_m\sin^{2m}\eta,
\\
\braket{(J,J),2J,0|j_{\xi}(\eta,\xi,\phi)|(J,J),2J,0}
=0,
\end{gathered}
\end{equation}
where $a_m$ and $b_m$ are arbitrary theory dependent real coefficients, subject to the constraint $\sum_m b_m=0$. Then the EFT gives
\begin{equation}
a_m=\begin{cases}
\frac{Q}{2\pi^2}, &\text{if } m=0,\\
0, & \text{if } 1\leq m\ll Q^{1/3};
\end{cases}
\qquad\qquad
b_m=\begin{cases}
\frac{3 Q^{2/3}}{8 \pi^2 \alpha}, &\text{if } m=0,\\
0, & \text{if } 1\leq m\ll Q^{1/3}.
\end{cases}
\end{equation}
Predictions are made only for $m\ll Q^{1/3}$ since the EFT breaks for distances of order of the inverse cutoff \eqref{eqDualCutoff} from the vortex, which lies at $\eta=\pi/2$. \footnote{To appreciate this, it is useful to write $\sin^{2m}\eta\approx \exp\left(-m\,\delta\eta^2\right)$ for $m\gg 1$ and $\delta\eta=\pi/2-\eta\ll 1$, which is exponentially suppressed away from the vortex core for $m\gtrsim Q^{1/3}$.}

A similar analysis can be done for the vortex crystal states in \eqref{eqVortexCrystal1} and \eqref{eqVortexCrystal2}. Consider first the traceless symmetric case $ Q\ll J_{34}\ll Q^{4/3}$ and $J_{12}=0$. Using \eqref{eqDensity1}, eq. \eqref{eqCorrelators} reads
\begin{equation}
\langle j_0\rangle= \frac{Q}{2\pi^2 R^3},\qquad
\langle j_{\phi}\rangle=\frac{e^2}{2\pi^2 R}\frac{J}{Q}
\cos^2\eta,\qquad
\langle j_{\xi}\rangle=0.
\end{equation}
This expression holds on scales larger than the vortex separation $\sim 1/\sqrt{\mJ}\sim \sqrt{Q/J}$, on which the continuous approximation \eqref{eqDensity1} can be used. It is then convenient to rewrite eq. \eqref{eqCorrelatorTTCFT} in Fourier basis
\begin{equation}\label{eqCorrelatorTTCFTfourier}
\begin{gathered}
\braket{(J,J),2J,0|j_{0}(\eta,\xi,\phi)|(J,J),2J,0}
=R^{-3}\sum_{m=0}^{2J}\tilde{a}_m\cos\left(2 m\,\eta\right),
\\ 
\braket{(J,J),2J,0|j_{\phi}(\eta,\xi,\phi)|(J,J),2J,0}
=R^{-3}\sum_{m=0}^{2J}\tilde{b}_m\cos\left(2 m\,\eta\right).
\end{gathered}
\end{equation}
Cutting off the sums at $m\ll\sqrt{\mJ}$, we obtain the following predictions
\begin{equation}
\tilde{a}_m=\begin{cases}
\frac{Q}{2\pi^2}, &\text{if } m=0,\\
0, & \text{if } 1\leq m\ll \sqrt{J/Q};
\end{cases}
\qquad\qquad
\tilde{b}_m=\begin{cases}
\frac{3 }{8 \pi ^2 \alpha }\frac{J_{34}}{Q^{1/3}}, &\text{if } m=0,1,\\
0, & \text{if } 2\leq m\ll \sqrt{J/Q}.
\end{cases}
\end{equation}
Analogously, for the state \eqref{eqVortexCrystal2} with $ Q\ll J_{12}, J_{34}\ll Q^{4/3}$, the EFT gives
\begin{equation}
\langle j_0\rangle= \frac{Q}{2\pi^2 R^3},\qquad
\langle j_{\phi}\rangle=\frac{e^2}{2\pi^2 R}\frac{J_{34}}{Q}
\cos^2\eta,\qquad
\langle j_{\xi}\rangle=
\frac{e^2}{2\pi^2 R}\frac{J_{12}}{Q}
\sin^2\eta.
\end{equation}
Without loss of generality, we assume $J_{12}\leq J_{34}$. The three-point function of a spin-1 conserved operator in a mixed symmetric state 
$\ket{(J,\bar{J}),J_{34},J_{12}}$, where $J_{34}$ and $J_{12}$ are related to $(J,\bar{J})$ as in \eqref{eqSummaryJJ}, can be conveniently written as \cite{Spinning3pt_Denis,Spinning4pt_Denis}:
\begin{equation}\label{eqCorrelatorMSCFT}
\begin{gathered}
\braket{(J,\bar{J}),J_{34},J_{12}|j_{0}(\eta,\xi,\phi)|(J,\bar{J}),J_{34},J_{12}}
=R^{-3}\sum_{m=0}^{2|J-\bar{J}|}a_m\cos(2m\,\eta),
\\ 
\braket{(J,\bar{J}),J_{34},J_{12}|j_{\phi}(\eta,\xi,\phi)|(J,\bar{J}),J_{34},J_{12}}
=R^{-3}\sum_{m=0}^{2|J-\bar{J}|+1}b_m\cos(2m\,\eta),
\\
\braket{(J,\bar{J}),J_{34},J_{12}|j_{\xi}(\eta,\xi,\phi)|(J,\bar{J}),J_{34},J_{12}}
=R^{-3}\sum_{m=0}^{2|J-\bar{J}|+1}c_m\cos(2m\,\eta).
\end{gathered}
\end{equation}
Here $a_m,b_m$ and $c_m$ are real coefficients, which satisfy the constraints
$\sum_m (-1)^m b_m=\sum_m c_m=0$ and $b_{2J+1}=-c_{2J+1}$. We then obtain the following results for the OPE coefficients:
\begin{equation}
\begin{gathered}
a_m=\begin{cases}
\frac{Q}{2\pi^2}, &\text{if } m=0,\\
0, & \text{if } 1\leq m\ll \sqrt{J_{12}/Q};
\end{cases}
\\
b_m=\begin{cases}
\frac{3 }{8 \pi ^2 \alpha }\frac{J_{34}}{Q^{1/3}}, &\text{if } m=0,1,\\
0, & \text{if } 2\leq m\ll \sqrt{J_{12}/Q};
\end{cases}
\quad\;
c_m=\begin{cases}
\frac{(-1)^m3 }{8 \pi ^2 \alpha }\frac{J_{12}}{Q^{1/3}}, &\text{if } m=0,1,\\
0, & \text{if } 2\leq m\ll \sqrt{J_{12}/Q}.
\end{cases}
\end{gathered}
\end{equation}

\section{Vortices in arbitrary dimensions}\label{secGeneralD}

Based on the considerations so far, as well as on the previous results of \cite{Cuomo}, it is not hard to understand the qualitative feature of the vortex EFT in higher spacetime dimensions. We give some brief comments here for completeness. We focus on the derivation of the scaling dimensions for traceless symmetric operators.

We first need to construct the dual of the $d+1$ dimensional Lagrangian \eqref{eqActionSuperfluid} in terms of a $d-1$ form gauge field $A$. Proceeding as in sec. \ref{secDualGaugeField}, this reads
\begin{equation}\label{eqDualActionD}
\mL=-\kappa \left|H\cdot H\right|^{\frac{d+1}{2d}},\qquad
H= d A.
\end{equation}
As in \eqref{eqCurrent}, the gauge and the scalar description are related by $\ast H\propto j$, where $\ast$ stands for the Hodge dual. The action \eqref{eqDualActionD}
can be expanded to quadratic order in terms of a non-propagating \emph{hydrophoton} $d-2$ gauge form and a longitudinal vector corresponding to the phonon.

Vortices are $d-1$ membranes which couple to the gauge field $A$ through a Kalb Ramond like interaction. Calling $X^\mu_p(\bar{\sigma})$ their line elements, 
where $\bar{\sigma}=(\tau,\sigma_1,\ldots)$ parametrizes the membrane coordinates, this coupling reads
\begin{equation}\label{eqKalbRamondD}
S_{KR}=-\sum_p\lambda_p \int d^{d-1}\bar{\sigma}A_{\mu_1\mu_2\ldots\mu_{d-1}}
\pd_{\tau}X_p^{\mu_1}
\pd_{\sigma_1}X_p^{\mu_2}\ldots\pd_{\sigma_{d-2}}X_p^{\mu_{d-1}}.
\end{equation}
One can similarly write the Nambu-Goto like action for the membrane \cite{MoninWheel}; we do not report here the expression since its detailed form will not be needed in the following. 

One can now proceed as in sec. \ref{secVortexEFT}. From the energy momentum tensor, one finds that the leading contribution to the vortex energy comes from the hydrophoton gauge field. Generalizing eq.s \eqref{eqAngularMom3d} and \eqref{eqAngularMomArea}, the angular momentum is proportional to the volume enclosed by the vortex in embedding coordinates. 

For $Q^{1/d}\ll J
\leq Q$, the minimal energy state corresponds to a single spherical vortex in embedding space. The leading contribution to the vortex energy arises from the running of the tension, induced by the hydrophoton contribution to the self-energy as in \eqref{eqEnergy1VortexPre}.
This can be computed using a flat space approximation for the gauge field Green function and a UV hard cutoff $\Lambda\sim Q^{1/d}/R$ to regulate the result:
\begin{equation}\label{eqJQenergyinD}
\Delta=\Delta_0(Q)+
\frac{d}{2 \alpha  (d+1)}
J^{\frac{d-2}{d-1}} Q^{\frac{1}{d(d-1)}} \log\left(J /Q^{\frac{1}{d}}\right),\qquad Q^{1/d}\ll J\leq Q,
\end{equation}
where $\Delta_0(Q)$ is given by \eqref{eqDeltaQdDimensions}. We expect $d$ dependent corrections of order $ J^{\frac{d-2}{d-1}} Q^{\frac{1}{d(d-1)}}$  to \eqref{eqJQenergyinD}, similarly to \eqref{eqDelta1Vortex}; these contributions however will not be logarithmically enhanced by the cutoff.

As in section \ref{secMultiVortices}, for $Q\ll J\ll Q^{\frac{d+1}{d}}$ we can identify the minimal energy state as a vortex crystal. Following the same steps which lead to \eqref{eqVortexCrystal1}, we find the energy of this state:
\begin{equation}\label{eqJbiggerQenergyinD}
\Delta
=\Delta_0(Q)+
\frac{d}{4 \alpha }
\frac{J^2}{ \,Q^{\frac{d+1}{d}}}, \qquad Q\ll J\ll Q^{\frac{d+1}{d}}.
\end{equation}
Eq.s \eqref{eqJQenergyinD} and \eqref{eqJbiggerQenergyinD} match the results obtained in \cite{Cuomo} and in this paper for $d=2,3$.

\section{Conclusions and future directions}\label{secConclusions}

Condensed matter phases often admit a simple \emph{effective} description \cite{Nicolis_Zoology}. In CFTs, one can take advantage of this using the state/operator correspondence to study CFT data at large quantum numbers. In this work, these ideas were used to compute the scaling dimensions of operators of large internal charge and spin in a $U(1)$ invariant CFT$_4$. The results obtained for traceless symmetric operators can be seen as a generalization of those obtained in CFT$_3$ \cite{Cuomo}; however the study of operators in mixed symmetric representations explored qualitatively distinct regimes, such as Kelvin wave propagation on a string (sec. \ref{secKelvinWaves}). We also provided predictions for correlators of the $U(1)$ current in between vortex states in sec. \ref{secCorrelators} and generalized the predictions for the scaling dimensions of traceless symmetric operators to arbitrary spacetime dimensions in eq.s \eqref{eqJQenergyinD} and \eqref{eqJbiggerQenergyinD}.

The most direct extension of this work would be a detailed analysis of higher order corrections, both in three and four dimensions. In particular, a refinement of the continuum approximation used in sec. \ref{secMultiVortices} might allow for the study of collective excitations in the vortex crystal phase, corresponding to an unexplored class of universal CFT operators, possibly similar to the Tkachenko mode studied in \cite{Moroz1,Moroz2}.

Superfluids and vortices are ubiquitous in physics. In the high energy physics literature, they are mostly studied in connection with the large density phase of  QCD \cite{Rajagopal:2000wf}. Neglecting the masses of the three lightest quarks, the superfluid modes are associated with the breaking of the baryon number $U(1)_B$, the strangeness $U(1)_S$ or the approximate axial symmetry\footnote{At high density, the effect of instantons breaking the $U(1)_A$ are suppressed \cite{Son:2000fh}.} $U(1)_A$. All other degrees of freedom are generically gapped but for a linear combination of the eighth gluon and the photon. For all of the superfluid modes vortices may arise \cite{StringsQCD} and they are indeed expected to exist in neutron stars
\cite{Kaplan:2001hh}. The physics of the $U(1)_B$ and the $U(1)_S$ vortices, which are electrically neutral, is analogous to the one studied in this paper and in general in \cite{Vortices}. The physics of the $U(1)_A$ vortices however entails some new features. Indeed the slight breaking of the symmetry implies the existence of (very fuzzy) domain walls surrounded by the string \cite{PhysRevLett.48.1867}. Furthermore, the $U(1)_A$ Goldstone may interact with the gauge fields via a Wess-Zumino-Witten term. Perhaps, some of these phenomena may be analyzed extending the EFT formalism of \cite{Vortices} used in this paper (see also \cite{Komargodski:2018odf} for interesting ideas in this direction).

Within the exploration of the superfluid phase of CFTs \cite{Hellerman,MoninCFT,Bern1,Bern2,Bern3,Bern4,Bern5,Bern6,CSLargeQ}, the non-Abelian case still leaves some open questions. As argued in \cite{MoninCFT}, the basic prediction for the scaling dimension of the lightest charged operator is insensitive to the non-Abelian nature of the symmetry group, which instead manifests itself via the existence of the so-called \emph{gapped Goldstones} \cite{Nicolis_Theorem,Nicolis_More}. Massive Goldstones are crucially
needed to close the current algebra of the non-Abelian group, like standard Goldstones, but at the same time have a fixed gap of order cutoff dictated by the symmetry. Their role in the large charge sector of CFTs and in more general finite density QFTs deserves further investigation.

Most of the existing results for large charge operators in CFTs are derived under the assumption that the CFT admits a superfluid phase. It is hence important to check whether this assumption applies in known theories. So far, most computations focussed on the prediction for the scaling dimension of the lightest charged operator. This has been verified in Monte-Carlo simulations
of the $O(2)$ \cite{LargeQMonteCarlo1} and $O(4)$ \cite{LargeQMonteCarlo2} model and perturbatively for Monopole operators \cite{Monopoles1,Monopoles2,Monopoles3,Pufu1,Pufu3}, to order $\mO(N^0)$ in the $\mathds{C}P^N$ model \cite{Anton} and at leading order in the number of flavors  in $QED_3$ and the gauged Gross-Neveu model \cite{William}.
Relatedly, large charge states have been studied in AdS/CFT in the context of \emph{holographic superconductors} \cite{Holography1,Holography2,Holography5}. We are currently addressing 
similar questions within the $\varepsilon$-expansion \cite{CuomoEpsilonInProgress},
which allows for extensive checks of the EFT predictions. Perhaps, the techniques explored in these works might also find application in a different context, such as the study of processes with many external legs within the Standard Model \cite{RubakovMany,SonMany,RattazziTrilinear,LutyTrilinear}.

Despite their generality, superfluids are not the only possible description for large charge states in CFT. For instance, if one assumes the charge $Q$ to be unbroken, a natural phase\footnote{If this is consistent with conformal invariance \cite{RothsteinDIHC,RothsteinFermiLiquid}.} which might describe the CFT is a Fermi liquid
\cite{Polchinski_Fermi}.
Different descriptions are also possible and are expected to apply in the presence of moduli spaces, which naturally allow for other light degrees of freedom different than the Goldstone mode. This is the case for free massless theories and $\mathcal{N}=2$ superconformal field theories, where the large $R$-charge expansion is organized differently \cite{SUSY1,SUSY2,SUSY3}.
In \cite{HellermanO41,HellermanO42} the possibility of a semiclassical but inhomogeneous phase in the $O(4)$ model was explored.

In \cite{BootstrapLargeQ},
the question of how to find solutions to the crossing equations at large charge was addressed in connection with the existence of a \emph{macroscopic limit} of correlators \cite{ETH_Dymarsky}. However, it remains an open question whether one can relate explicitly the large charge sector of CFTs with the spectrum of light operators. Perhaps relatedly, a bootstrap analysis recently connected
large scaling dimensions tails of weighted spectral density of primary operators with light operators exchanged in the dual channel
\cite{LargeD_bootstrap}.

Finally, the predictions of analyticity and the existence of a perturbative expansion for large internal quantum numbers are reminiscent of the bootstrap results for large spin operators \cite{LargeS1,LargeS2,LargeS_Mat1,LargeS_Alday1,LargeS_Alday2,
LargeS_Alday3,LargeS_Alday5,LargeS_Alday6,
LargeS_Kaviraj1,LargeS_Kaviraj2,LargeS_Ising}. The physical picture behind those results is particularly clear in the dual AdS space\footnote{A similar picture was proposed in \cite{LargeS0}, with no reference to the gravity dual.} \cite{LargeS2,LargeS_Mat1}, where double-trace operators correspond to 
two widely separated objects. Because of the AdS geometry, these only interact weakly via the exchange of highly off-shell modes. A universal EFT description might exist in this case as well.

\subsection*{Acknowledgements}

I thank Anton de la Fuente, Alexander Monin, and Riccardo Rattazzi for useful discussions and valuable comments on the draft. I would also like to acknowledge the reviewer for insightful comments. My work is partially supported by the Swiss National Science Foundation under contract 200020-169696 and through the National Center of Competence in
Research SwissMAP.

\appendix

\section{Vortices with half integer spin in 2+1 dimensions}\label{spinV}

In this section, we provide some details on vortices with half-integer spin, discussed in sec. \ref{secVortices3D}. To this aim we modify the action \eqref{eqAction3dVortices} following the approach of \cite{spin1,spin2}. 

In a parity invariant setting, an half integer spin particle at rest in $2+1$ dimensions can be in two different states. To add such a discrete multiplicity of states it is natural to use a Grassmanian field living on the worldline, transforming covariantly under the unbroken rotations\footnote{The worldline action breaks spontaneously boosts, see appnedix \ref{AppendixCoset}.}. To do so in a covariant way, we consider a Grassmanian $3$-vector $\xi^\mu_p(\tau)$ orthogonal to the particle velocity: $\xi_p^\mu g_{\mu\nu}\dot{X}^\nu_p=0$. Here $p$ labels the different vortices. Neglecting interactions for the moment, the free reparametrization invariant action for one such variable reads:
\begin{equation}\label{Lag1}
S^{(p)}_{\xi}\big\vert_{free}=-\frac{i}{2}\int d\tau\left[\xi^\nu_p g_{\mu\nu}\left(\frac{D}{D\tau}\xi^\mu_p\right)+
\lambda_p\left(\dot{X}^\mu_p g_{\mu\nu}\xi^\nu_p\right)\right],
\end{equation}
where $\lambda_p$ is a Grassmanian Lagrange multiplier. 

The Lagrangian in eq. \eqref{Lag1} was first studied in \cite{spin1} \footnote{The authors of \cite{spin1} consider an additional redundant variable $\xi_5$; eq. \eqref{Lag1} coincides with their eq. (3.3) in the gauge $\xi_5=0$.}, where it was proven to correctly describe a parity-invariant Dirac particle. We shall not repeat the full analysis here, but it is useful to build some intuition focussing the non-relativistic limit of \eqref{Lag1} in flat space. Working in the physical gauge $X^0_p=t$, the constraint $\xi_p^\mu \eta_{\mu\nu}\dot{X}^\nu_p=0$ implies $\xi^0_p\simeq 0$ and we obtain
\begin{equation}
S^{(p)}_\xi\big\vert_{free}\simeq
\frac{i}{2}\int d\tau \xi^i_p \delta_{ij}\dot{\xi}^j_p\,.
\end{equation}
Quantizing the constrained system we get $\{\xi^i_p,\xi^j_p\}=-\delta^{ij}$ \cite{spin2}, from which we can identify $\xi^i_p=i\sigma^i/\sqrt{2}$ ($i=1,2$) and the spin is the conserved operator $s^{12}_p= -i/2[\xi_p^1,\xi_p^2]=-\sigma^3/2$. The two eigenstates of $s^{12}_p$ describe the possible spin orientations for a particle at rest. 

Let us now consider the possible interaction terms which can be introduced in \eqref{Lag1}. Considering that the action shoud be quadratic in $\xi^\mu_p$, that classically $\xi_p\cdot\xi_p=0$ and that we work at leading order in the particle velocities, the leading interaction term is given by
\begin{equation}
\frac{i}{2}\frac{g_p}{\sqrt{2}}\int d\tau \sqrt{\dot{X}^\mu_p g_{\mu\nu}\dot{X}^\nu_p}\frac{F_{\mu\nu}\xi^\mu_p\xi^\nu_p}{\sqrt{F/\sqrt{2}}}\, ,
\end{equation}
which can be interpreted as the relativistic version of the Pauli interaction \cite{spin0}. The powers of $F$ are dictated by Weyl invariance and the dimensionless coupling $g_p$ can be interpreted as the magnetic moment of the particle. We then conclude that, in order to describe fermionic vortices with half-integer spin, we need to add to the action \eqref{eqAction3dVortices} the following term
\begin{equation}\label{Lag}
\sum_p S^{(p)}_\xi=-\frac{i}{2}\sum_p\int d\tau\left[\xi^\nu_p g_{\mu\nu}\left(\frac{D}{D\tau}\xi^\mu_p\right)+
\lambda_p\left(\dot{X}^\mu_p g_{\mu\nu}\xi^\nu_p\right)
-\frac{g_p}{\sqrt{2}}
 \sqrt{\dot{X}^\mu_p g_{\mu\nu}\dot{X}^\nu_p}\frac{F_{\mu\nu}\xi^\mu_p\xi^\nu_p}{\sqrt{F/\sqrt{2}}}\right]\,.
\end{equation}

To study the model given by \eqref{eqAction3dVortices}+\eqref{Lag}, it is useful to integrate out the Coulomb field $A_0(x)$ and notice that to leading order in the velocities $\xi^0_p\simeq0$. It is further convenient to introduce a Grassmanian vector in embedding space as
\begin{equation}
\xi_p^a=\frac{d X^a_p}{d x^i_p}\xi^i_p,\qquad
\vec{X}_p\cdot\vec{\xi}_p=0\,,
\end{equation}
where $\vec{X}_p$ is the particle position in embedding coordinates. Then the effective action for the vortices reads
\begin{multline}\label{LagFinal}
S\simeq\int dt \left\{\sum_p\vec{A}\cdot\dot{\vec{X}}_p+\frac{e^2}{8\pi}\sum_{p,r}q_pq_r\log\left(\vec{X}_p-\vec{X}_r\right)^2\right.\\
\left.+\frac{i}{2}\sum_p
\left[\vec{\xi}_p\cdot\dot{\vec{\xi}}_p-\frac{g_p}{\sqrt{2}}\sqrt{B}\epsilon_{abc}X^a_p\xi^b_p\xi^c_p\right]\right\}
\end{multline}
where $\vec{A}$ is the potential for a magnetic monopole \cite{Wu:1976ge}. We neglected fluctuations of $A_i$ and the coupling between the electric field and $\vec{\xi}_p$ to leading order. The first two terms were studied in \cite{Cuomo}. Each particle gives both an orbital and a spin contribution to the angular momentum:
\begin{equation}
\vec{J}=\sum_p\left(\vec{L}_p+\vec{s}_p\right),\qquad
L^a_p=-\frac{Q}{2}q_pX^a_p,\qquad s^a_p=-\frac{i}{2}\epsilon_{abc}\xi^b_p \xi^c_p,
\end{equation}
where, upon quantization, $[L^a_p,L^b_p]=i\epsilon_{abc}L^c_p$ and $[s^a_p,s^b_p]=i\epsilon_{abc}s^c_p$, with $\vec{s}_p\cdot\vec{s}_p=3/4$. Notice that only $\vec{J}$ is conserved, while the orbital and the spin part alone are not, due to the interaction of the spin with the magnetic monopole field. The Hamiltonian is
\begin{equation}
H=\text{const}+\frac{e^2}{4\pi}\log\vec{L}^2+\sum_p\frac{g_p}{q_p}\frac{\vec{L}_p\cdot\vec{s}_p}{R\sqrt{Q}}\,.
\end{equation}

Consider finally a semiclassical state made of a unit charge vortex-antivortex pair of the kind considered around eq. \eqref{eqDelta3D2Vortices}. We assume that both vortices have the same magnetic moment $g>0$. For such a state the spin contribution to the angular momentum is negligble: $\vec{J}\simeq \vec{L}=Q\, \Delta\vec{X}$. To compute the energy of this state we use $\langle\vec{L}_p\cdot\vec{s}_p\rangle\simeq \vec{L}^{class.}_p\cdot\langle\vec{s}_p\rangle$, where $\vec{L}_p^{class.}$ is obtained solving the classical equations of motion at fixed angular momentum. By properly choosing the spin orientation of the particles to minimize the energy, we obtain
\begin{equation}
\Delta=\Delta_0(Q)+\frac{\sqrt{Q}}{3\alpha}\log\frac{J}{\sqrt{Q}}+2\tilde{\gamma}\sqrt{Q}-\frac{g}{2}\frac{J}{\sqrt{Q}},\qquad \sqrt{Q}\ll J\leq Q\,,
\end{equation}
which differs by the prediction for two spinless vortices in eq. \eqref{eqDelta3D2Vortices} only by the last term.

\section{Photon propagator on the sphere}\label{AppProp}

Here we obtain the photon propagator on a d dimensional sphere following the simple method\footnote{A similar derivation in $4d$ de Sitter can be found in \cite{propdS}.} of \cite{propAdS1,propAdS2}. In this section we set $R=1$. 

Consider the action of a massless vector field coupled to a conserved current $J^\mu$ (in Euclidean signature):
\begin{equation}
S=\int d^d x\sqrt{g}\left(\frac{1}{4}f_{\mu\nu}f^{\mu\nu}-a_{\mu}J^\mu\right),\qquad
f_{\mu\nu}=\pd_\mu a_\nu-\pd_\nu a_\mu.
\end{equation}
The gauge field on the equations of motion is given by
\begin{equation}\label{eqAppPhoton}
a_\mu(x)=\int d^dx'\sqrt{g'} G_{\mu\nu'}(x,x')J^{\nu'}(x'),
\end{equation}
where $G_{\mu\nu'}(x,x')$ satisfies the equation
\begin{equation}\label{eqAppPropagatorEq}
\nabla^\mu\left(\pd_\mu G_{\nu\nu'}(x,x')
-\pd_\nu G_{\mu\nu'}(x,x')\right)=-g_{\nu\nu'}(x)\frac{\delta(x-x')}{\sqrt{g'}}+
\pd_{\nu'}\Lambda_\nu(x,x').
\end{equation}
Here $\Lambda_\nu$ is a pure gauge term which drops from physical observables; primed and unmprimed indices refer, respectively, to the points $x$ and $x'$.

Let us define the following biscalar
\begin{equation}
u=\frac{1}{2}(X-X')^2,
\end{equation}
where $(X-X')^2$ is the chordal distance in embedding coordinates. Given the 
isometries of the sphere, it is possible to parametrize the propagator as
\begin{equation}\label{eqAppProp}
G_{\nu\nu'}(x,x')=-\left(\pd_\nu\pd_{\nu'}u\right)F(u)+\pd_\nu\pd_{\nu'}S(u).
\end{equation}
The last term is gauge dependent and drops from eq. \eqref{eqAppPhoton}.

The following properties hold:
\begin{enumerate}
\item $\nabla^\mu \pd_\mu u=d(1-u)$ ,
\item $g^{\mu\nu}\pd_\mu u\pd_\nu u=u(2-u)$,
\item $\nabla_\mu \pd_\nu u=g_{\mu\nu}(1-u)$,
\item $(\nabla^\mu u)(\nabla_\mu\pd_{\nu'}u)=(1-u)\pd_{\nu'}u$,
\item $(\nabla^\mu u)(\nabla_\mu\pd_{\nu}\pd_{\nu'}u)=-\pd_\nu u\pd_{\nu'}u$.
\end{enumerate}
These can be explicitly verified in stereographic coordinates, for instance.
It follows
\begin{multline}
\nabla^\mu\left(\pd_\mu G_{\nu\nu'}(x,x')
-\pd_\nu G_{\mu\nu'}(x,x')\right)=
-\left(\pd_\nu\pd_{\nu'}u\right)\left[u(2-u)F''+(d-1)(1-u)F'\right]\\
+\left(\pd_\nu u\pd_{\nu'}u\right)\left[(1-u)F''+(1-d)F'\right].
\end{multline}
By symmetry, we can write $\Lambda_\nu(x,x')=(\pd_\nu u)\Lambda(u)$. Then for $x\neq x'$ \eqref{eqAppPropagatorEq} gives two equations:
\begin{align}
&u(2-u)F''+(d-1)(1-u)F'=-\Lambda,\\
&(1-u)F''-(d-1)F'=\Lambda'.
\end{align}
We can integrate the second and plug the result in the first to obtain
\begin{equation}
(2-u)u F''(u)+d(1-u)F'(u)-(d-2)F(u)=0.
\end{equation}
This is just Klein Gordon equation for a scalar field of mass $m^2=d-2$ on $S^d$. The solution is fixed requiring a power low singularity for $u\rightarrow 0$ and regularity at the antipodal point $u\rightarrow 2$ \cite{propJacobson}:
\begin{equation}\label{eqAppPropF}
F(u)=\frac{\Gamma(d-2)}{(4\pi)^{\frac{d}{2}}\Gamma\left(\frac{d}{2}\right)}
\,_2 F_1\left(1,d-2;\frac{d}{2};1-\frac{u}{2}\right),\qquad\qquad d>2.
\end{equation}
The normalization is determined matching the short distance limit
with the flat space propagator. Plugging in \eqref{eqAppProp}, we get eq. \eqref{eqPhotonProp} in the main text.

\section{Vortex energy in dimensional regularization}\label{AppMagnetostaticEnergy}

To regulate the computation of the magnetostatic energy, it is convenient to work in $d+1$ spacetime dimensions. It is natural to modify the Lagrangian \eqref{eqActionPureGauge} in a way which preserves Weyl invariance:
\begin{equation}\label{eqAppActionPureGaugeD}
\mL=-\kappa H^{(d+1)/3}.
\end{equation}
The definition of $H$ in terms of the two-form field $A_{\mu\nu}$ here is unchanged. Notice that working in arbitrary $d$ with a 2-form field we loose the duality with a shift invariant scalar, hence this regularization breaks $U(1)$ invariance at intermediate steps in the calculation\footnote{Conversely, a cutoff approach as in \cite{Cuomo} breaks Weyl invariance.}. A more responsible approach might be to promote $A_{\mu\nu}$ to a $d-2$ form field, preserving Weyl invariance of the action. An investigation of this issue might be helpful in expanding the result of this paper to subleading orders.

Expanding the action \eqref{eqAppActionPureGaugeD} to quadratic order gives
\begin{equation} 
\mL_{fluct}=\frac{1}{4 e^2(d)}f_{ij}f^{ij}+\frac{1}{2 e^2(d)}\left[\dot{b}^i\dot{b}_i-\frac{(d-3)}{3}(\nabla_i b^i)^2\right],
\end{equation}
where we defined the electric coupling in $d$ space dimensions as
\begin{equation}
e^2(d)=\frac{\left(\sqrt{6}B\right)^{2-\frac{d+1}{3}}}{2(d+1)\kappa}
=e^2\left[1
-(d-3)\left( \log B^{\frac{1}{3}}+\frac{1}{4}+\frac{1}{6}\log 6\right)
+\mO\left((d-3)^2\right)\right].
\end{equation}
The NG action discussed in section \ref{secStringVortex} is unchanged in $d$ dimensions.

\subsection{Vortex ring self-energy}\label{App_self_vortex}

Consider a single vortex moving on a trajectory given by
\eqref{eqSingleVortexConfig}. We want to compute the self-energy contribution due to the hydrophoton, i.e. the second term in eq. \eqref{eqEnergy1VortexPrePre}. In Hopf coordinates \eqref{eqHopf} and in dimensional regularization, it reads:
\begin{equation}
\begin{split}
E_{mag}&=\frac{ e^2(d)}{2}R^2\iint d\xi d\xi'  G_{\xi\xi}\left((\eta,\xi,\phi);
(\eta,\xi',\phi)\right) 
= \pi  R e^2(d)R^{3-d}I(r,d),
\end{split}
\end{equation}
where we isolated the integral
\begin{equation}\label{eqAppInt1}
I(r,d)=r^2
\frac{ \Gamma (d-2)}{(4\pi)^{d/2}\Gamma \left(\frac{d}{2}\right)}
\int_0^{2\pi}d\xi\cos (\xi ) \, _2F_1\left(1,d-2;\frac{d}{2};1-\frac{1}{2} r^2 (1-\cos \xi )\right).
\end{equation}
In $d=3$, the integral is logarithmically divergent for $\xi\rightarrow 0$, corresponding to the interaction of an infinitesimal line element with itself.

Setting $\frac{1}{2} (1-\cos\xi )=y$ in \eqref{eqAppInt1}, we get
\begin{equation}
I(r,d)=2r^2
\frac{ \Gamma (d-2)}{(4\pi)^{d/2}\Gamma \left(\frac{d}{2}\right)}\int_0^1 dy \frac{(1-2 y) }{\sqrt{(1-y) y}}\, _2F_1\left(1,d-2;\frac{d}{2};1-r^2 y\right).
\end{equation}
The divergent part comes from the first term in the expansion of the hypergeometric function when the argument goes to one:
\begin{equation}\label{eqApp2F1expansion}
\, _2F_1\left(a,b;c;1-z\right)
\xrightarrow{z\rightarrow 0}\frac{1}{z^{a+b-c}}
\frac{\Gamma (c) \Gamma (a+b-c)}{\Gamma (a) \Gamma (b)}
 ,\qquad a+b> c.
\end{equation}
We separate explicitly this contribution and recast the integral as:
\begin{equation}
I(r,d)=I_{div}(r,d)+I_{reg}(r,d),
\end{equation}
where the divergent piece is
\begin{equation}
\begin{split}
I_{div}(r,d)&=2r^2\frac{  \Gamma \left(\frac{d}{2}-1\right)}{(4\pi)^{d/2}}
\int_0^1 dy
\frac{(1-2 y) }{\sqrt{(1-y) y} }\left(\frac{1}{r^2 y}\right)^{\frac{d-2}{2}}\\
&=\frac{r}{2 \pi  (3-d)}+\frac{r \left[\log \left(4 \pi  r^2\right)-\gamma_E -
2 \psi \left(3/2\right)\right]}{4 \pi }
+\mO\left((3-d)\right),
\end{split}
\end{equation}
and the regular part can be evaluated directly in $d=3$, where it reads
\begin{equation}
I_{reg}(r)\equiv I_{reg}(r,3)=
\frac{r}{2 \pi ^2}
 \int_0^1 dy
\frac{(1-2 y) }{ y \sqrt{1-y}  }\left[\frac{\arcsin\left(\sqrt{1-r^2 y}\right)}{\sqrt{1-r^2 y}}-\frac{\pi}{2} \right].
\end{equation}
To compute the latter, it is convenient to use the following expansion
\begin{equation}
\frac{\arcsin\left(\sqrt{1-x^2}\right)}{\sqrt{1-x^2}}=
\sum _{m=0}^{\infty } \frac{-(-2)^{m+1} \Gamma \left(\frac{m+3}{2}\right)^2 }{(m+1)^2 }\frac{x^m}{m!},\qquad
0\leq x< 1.
\end{equation}
Interchanging sum and integral, the regular part gives
\begin{equation}
I_{reg}(r)=\frac{r^2}{4 \pi ^2} \sum_{m=1}^\infty
\frac{ (-1)^{m+1} 2 \pi(m-1) r^{m-1}}{m (m+1)}=
\frac{r}{\pi}-\frac{r}{2\pi}\log (r+1)-\frac{1}{\pi}\log (r+1).
\end{equation}
Collecting everything and adding the tension contribution, we arrive at \eqref{eqEnergy1VortexPre}.

\subsection{Kelvin waves frequency}\label{App_LinearKelvon}

The EOMs which derive from \eqref{eqSingleVortexActionSmall} give the oscillation frequency of Kelvin waves as
\begin{equation}
\begin{split}
\frac12B\omega_n &=
\gamma\frac{\pi  B^{2/3}}{R^2}   \left(n^2-1\right) +
\frac{2\pi e^2(d)R^{3-d}}{R^2}\delta\omega_n^I
\end{split},
\end{equation}
where the second term comes from the nonlocal piece of the action and is written in terms of the following integral:
\begin{multline}\label{eqAppKelvinEnergyInt}
\delta\omega_n^{I}=\frac12\int d\sigma\Big\{\left[
n^2\cos(n\sigma)-\cos\sigma\right]F(1-\cos\sigma)\\
+\left[\cos^2\sigma-
\cos(n\sigma)\cos\sigma\right]F'(1-\cos\sigma)\Big\}.
\end{multline}
Let us sketch the evaluation of \eqref{eqAppKelvinEnergyInt}.
Changing variables as before, we write $\delta\omega_n^{I}$ as the sum of the following two contributions:
\begin{equation}
I_1(n)=\frac{\Gamma(d-2)}{(4\pi)^{d/2}\Gamma(d/2)}\int_0^1 \frac{dy}{\sqrt{y(1-y)}}\left[
n^2T_n(1-2y)-(1-2y)\right]
\,_2 F_1(1,d-2;d/2;1-y),
\end{equation}
\begin{multline}\label{eqAppKelvinI2}
I_2(n)=\frac{\Gamma (d-1)/2}{ (4 \pi )^{d/2} \Gamma \left(d/2+1\right)}
\\
\times\int_0^1 \frac{dy(1-2y)}{\sqrt{y(1-y)}}\left[T_n(1-2y)-(1-2y)
\right]\,_2 F_1(2,d-1;d/2+1;1-y).
\end{multline}
Here $T_n(x)=\cos\left(n\arccos(x)\right)$ is a Chebyshev polynomial.
The divergent contributions are identified from the leading term of the Hypergeometric expansion \eqref{eqApp2F1expansion} and can be evaluated using 
\begin{equation}\label{eqUsefulIntegral}
\int _0^1 dy\frac{T_n(1-2 y)}{\sqrt{y (1-y)}}y^{m-\frac{1}{2}} =
\frac{\sqrt{\pi } \Gamma (m) \left(\frac{1}{2}-m\right)_n}{\Gamma \left(m+n+\frac{1}{2}\right)}.
\end{equation}
To evaluate the regular parts, we use the following results:
\begin{equation}\label{eqAppLn}
\begin{split}
\lambda_n&\equiv
\int_0^1 dy\frac{T_n(1-2y)}{\sqrt{y(1-y)}}\left[\frac{\arcsin\left(\sqrt{1-y}\right)}{\sqrt{y}\sqrt{1-y}}-\frac{\pi }{2\sqrt{y}}\right]\\
&=\frac{\pi}{2}   \left[
\psi\left(\frac{n}{2}+1\right)+2 \psi\left(n+\frac{1}{2}\right)-\psi\left(\frac{n+1}{2}\right)
-2 \psi\left(n+1\right)\right],
\end{split}
\end{equation}
\begin{equation}\label{eqAppRhon}
\begin{split}
\rho_n\equiv&
\int _0^1 dy
 \frac{1-2 y}{\sqrt{y (1-y)}}T_n(1-2 y) \left[
 \, _2F_1\left(2,2;\frac{5}{2};1-y\right)-\frac{3 \pi}{8y^{3/2}}
 +\frac{3\pi}{16 y^{1/2}}\right]
\\
=&\frac{3}{2} \pi  \left\{ \left(n^2+1\right) \left[\psi\left(\frac{n-1}{2}\right)-\psi\left(n-\frac{1}{2}\right)+\log 2\right]+\frac{ 4 n^4+6 n^2+3 n-1}{4 n^3-4 n^2-n+1}+\frac{3}{8}\right\}.
\end{split}
\end{equation}
Using \texttt{Mathematica} we computed these integrals explicitly for fixed integer values of $n$ and
identified their functional form; the result was then verified numerically and using the $n\rightarrow \infty$ asymptotic expansion of the results \eqref{eqAppLn} and \eqref{eqAppRhon}. This indeed can be obtained explicitly truncating the series expansion of the Hypergeometric functions in the integrals and using \eqref{eqUsefulIntegral}.
The regular contributions finally read
\begin{equation}
I_1^{reg}(n)=
\frac{1}{4\pi^2}\left(n^2\lambda_n-\lambda_1\right),
\end{equation}
\begin{equation}\label{eqAppKelvinI2reg}
\begin{split}
I_2^{reg}(n)&= \frac{1}{12\pi^2}(\rho_n-\rho_1)-\frac{1}{64\pi}\int _0^1 dy
 \frac{(1-2 y)\left[T_n(1-2 y) -(1-2 y)\right]}{y\sqrt{ (1-y)}}\\
&=\frac{1}{12\pi^2}(\rho_n-\rho_1)+ \frac{3 \psi\left(n+\frac{1}{2}\right)+3 \gamma_E -4+\log (64)
-\frac{6}{4 n^2-1}}{96 \pi }.
\end{split}
\end{equation}
The second contribution in \eqref{eqAppKelvinI2reg} arises since we subtracted the $\mO\left(1/\sqrt{y}\right)$ term in the expansion of the Hypergeometric function from the first piece, in order to apply \eqref{eqAppRhon}.
Collecting everything and expanding for $d\rightarrow 3$, we find the following remarkably simple result:
\begin{equation}
\delta\omega_n^I=\frac{n^2-1}{8\pi  (3-d)}
+\frac{\left(n^2-1\right) \left[\log\pi-2 \psi\left(\frac{n+1}{2}\right)-\gamma_E -1 \right]}{16 \pi}.
\end{equation}
Eq. \eqref{eqKelvonFreq} then follows.

\section{Nambu-Goto action from the coset construction}\label{AppendixCoset}

In \cite{MoninCFT}, it was argued that two charged scalar operators insertions in $d+1$ dimensions at $x=0$ and $x=\infty$ induce a specific symmetry breaking pattern for the leading trajectory in the path integral. A similar logic can be applied when the operators have also large spin $J$. As in the scalar case, translations $P_\mu$, special conformal transformations $K_\mu$ and dilatation $D$ are broken, with the combination $D+\mu Q$ left unbroken. Assuming the operator insertion to be polarized in the $(x_1,x_2)$ plane, the Lorentz generators $J_{1p}, \;J_{2p}$ with $p,q=0,3,\ldots$ must necessarily be broken.
A vortex corresponds to the regime where it is energetically favorable for the system to still be in an almost homogeneous state, rotations being broken by a localized region of size $1/j_0\sim R/Q^{1/d}$ in which the superfluid description breaks. This region naturally extends
from $0$ to $\infty$ along the directions orthogonal to the spin polarization, corresponding hence to a $d-1$ dimensional membrane. In this regime, $J_{12}$ parametrizes rotation around the vortex and it is thus unbroken. We then identify the symmetry breaking pattern corresponding to a vortex as:
\begin{equation}\label{eqAppSBvortex}
\begin{cases}
\bar{D}=D+\mu \,Q, J_{12},J_{pq} & \text{unbroken},\\
D,P_\mu,K_\mu,J_{mp} & \text{broken}.
\end{cases}
\end{equation}
where we introduced the set of indices $m,n=1,2$ and $p,q=0,3,\ldots$ . 

In order to apply the coset construction in a curved manifold, it is convenient to think in terms of the generators acting in a local chart, denoted $\{\widehat{D},\widehat{P}^\mu,\widehat{K}^\mu,\widehat{J}_{\mu\nu}\}$. These  are naturally associated with those acting on the plane considering the formal $R\rightarrow\infty$ limit on $\mathds{R}\times S^d$ \cite{MoninCFT}:
\begin{equation}\label{eqAppHattedGenerators}
\begin{gathered}
D=-R\widehat{P}_0,\qquad J_{ij}=\widehat{J}_{ij},\qquad J_{0i}=- R\widehat{P}_i,
\qquad P_0=\widehat{P}_0+\frac{\widehat{D}}{R}+\frac{\widehat{K}_0}{2R^2},\\
K_0=\frac{1}{2}\widehat{K}_0-R\widehat{D}+R^2\widehat{P}_0,\qquad
P_i=\widehat{P}_0+\frac{\widehat{J}_{0i}}{R}-\frac{\widehat{K}_i}{2R^2},
\qquad
K_i=\frac12 \widehat{K}_i+ R\widehat{J}_{0i}-R^2\widehat{P}_i.
\end{gathered}
\end{equation}
We then rewrite the symmetry breaking pattern \eqref{eqAppSBvortex} in terms of the hatted generators. Focussing on $2+1$ and $3+1$ dimensions, we get
\begin{equation}\label{eqAppSBvortexGauged}
2+1:
\begin{cases}
\widehat{\bar{P}}_0=\widehat{P}_0+\mu \,Q, \widehat{J}_{12} & \text{unbroken},\\
\widehat{P}_i,\widehat{J}_{0i},\widehat{K}_\mu,\widehat{Q} & \text{broken};
\end{cases}\qquad
3+1:
\begin{cases}
\widehat{\bar{P}}_p=\widehat{P}_p+\mu \delta_p^0\,Q, \widehat{J}_{12} & \text{unbroken},\\
\widehat{P}_m,\widehat{J}_{0i},\widehat{J}_{n3},\widehat{K}_\mu,\widehat{Q} & \text{broken}.
\end{cases}
\end{equation}
From \eqref{eqAppSBvortexGauged} we can construct the Nambu-Goto action  for the vortex via the coset construction \cite{CCWZ1,CCWZ2}, applied to the case of a membrane \cite{MoninWheel}.

\subsection{2+1 dimensions}

Following \cite{MoninCFT}, we gauge all spacetime symmetries and specify the manifold only at the end of computations. We henceforth do not consider special conformal transformations anymore and work with a covariant derivative in terms of three gauge connections:
\begin{equation}
D_\mu=\pd_\mu+i \tilde{e}^a_\mu P_a+\frac{i}{2}\omega^{ab}_\mu J_{ab}+i A_\mu D. 
\end{equation}
Inices $a,b=0,1,\ldots$ label the gauged Poincar\'e generators and should not be confused with spacetime indices $\mu,\nu,\ldots$ \cite{IvanovGravity1,IvanovGravity2}.
From \eqref{eqAppSBvortex}, the coset of a vortex line in $2+1$ dimensions is formally identical to the conformal superfluid one 
\begin{equation}
\Omega=e^{i y^a \bar{P}_a}e^{i\sigma D}e^{i\eta^i J_{0i}}e^{i\pi Q}=
e^{i y^a P_a}e^{i\sigma D}e^{i\eta^i J_{0i}}e^{i\chi Q},\qquad
\chi=\mu t+\pi,
\end{equation}
The Maurer-Cartan (MC) one form reads
\begin{equation}\label{eqAppMC1form3D}
\Omega^{-1}D_\mu \Omega
=i E^a_\mu\left(\bar{P}_a+\nabla_a\sigma D+\nabla_a\pi Q+\nabla_a\eta^i
J_{0i}+\frac12 \Omega^{ij}_a J_{ij}
\right);
\end{equation}
where
\begin{equation}
E^a_\mu=e^{-\sigma}e^b_\mu\Lambda_b^{\,a},\qquad
\nabla_a\pi=e^{\sigma}e^{\mu}_b\Lambda^b_{\,a}\pd_\mu\chi-\mu\delta_a^0,\qquad
\nabla_a\sigma=e^{\sigma}e^{\mu}_b\Lambda^b_{\,a}\left(\pd_\mu\sigma+A_\mu\right).
\end{equation}
Here $e^a_\mu$ transforms as a vierbein. We introduced the Lorentz matrix $(e^{-i\eta^i J_{0i}})_{\;b}^{a}=\Lambda_b^{\;a}$. The expressions of $\nabla_a\eta^i$ and $\Omega_a^{ij}$ are not needed here. 
One can also construct curvature invariants as
\begin{equation}\label{eqAppMCcurvature3D}
\Omega^{-1}[D_\mu,D_\nu]\Omega=
i E^a_\mu E^b_\nu\left(T^c_{ab}P_c+\frac12 R^{cd}_{ab}J_{cd}+
A_{ab} D\right).
\end{equation}
Explicit expressions for these can be found in \cite{MoninCFT}.
Finally, one needs to consider the projection of the MC one form onto the vortex world-line $x^\mu(\lambda)$ \cite{MoninWheel}:
\begin{equation}\label{eqAppMC1form3Dprojected}
\dot{x}^\mu \Omega^{-1}D_\mu \Omega
=i E \left(P_0+\nabla y^i P_i+\nabla\sigma D+\nabla\chi Q+\nabla\eta^i
J_{0i}+\frac12 \Omega^{ij} J_{ij}
\right),
\end{equation}
where
\begin{equation}
\begin{gathered}
E=\dot{x}^\mu e^{-\sigma}e^b_\mu\Lambda_b^{\,0},\qquad\quad
\nabla y^i=E^{-1}\dot{x}^\mu e^{-\sigma}e^b_\mu\Lambda_b^{\,i},
\\
\nabla\chi=E^{-1}\dot{x}^\mu\pd_\mu\chi,\qquad\quad
\nabla\sigma=E^{-1}\dot{x}^\mu\left(\pd_\mu\sigma+A_\mu\right).
\end{gathered}
\end{equation}
We can reduce the number of independent Goldstones setting to zero one or more of the invariants in \eqref{eqAppMC1form3D}, \eqref{eqAppMCcurvature3D} or \eqref{eqAppMC1form3Dprojected}. When an algebraic solution exists, these conditions are called Inverse Higgs Constraints (IHCs) \cite{IvanovIHC,LowIHC}. In this case, the same IHCs which lead to the superfluid action are imposed:
\begin{equation}\label{eqAppIHC1}
T^a_{bc}=0,\qquad \nabla_0\pi=0,\qquad\nabla_i\pi=0,\qquad
\nabla_a\sigma=0.
\end{equation}
The first is a \emph{torsion free} condition and selects the spin one-connection $\omega^{ab}_\mu$ compatible with the metric $\hat{g}_{\mu\nu}=e^{-2\sigma}g_{\mu\nu}$. The others are used to express $A_\mu,\sigma$ and $\eta^i$ in terms of the other fields:
\begin{equation}\label{eqAppIHC1sol}
A_\mu=-\pd_\mu\sigma,\qquad
\mu e^{-\sigma}=(e^{\mu a}e^\nu_a\pd_\mu\chi\pd_\nu\chi)^{1/2},\qquad 
\frac{\eta^i}{\eta}\tanh\eta=-\frac{e_i^\mu\pd_\mu\chi}{e_0^\mu\pd_\mu\chi},
\end{equation}
where $\eta\equiv\sqrt{\eta^i\eta^i}$ and $(\pd\chi)=(e^{a\mu}e_a^\nu\pd_\mu\chi\pd_\nu\chi)^{1/2}$.

We can now construct the leading order invariants in the world-line. Noticing that $\nabla\chi=\mu$ and $\nabla\sigma=0$, these are constructed out
of the \emph{einbein} $E$ and the covariant derivative $\nabla y^i$ as: 
\begin{equation}\label{eqAppInvariants3d}
\mu E=\dot{x}^\mu \pd_\mu\chi,\qquad
\nabla y^i\nabla y^i=1-
\frac{(\pd\chi)^2\dot{x}^\mu \dot{x}_\mu}{(\dot{x}^\mu \pd_\mu\chi)^2}.
\end{equation}
The most general NG action is then written in terms of an arbitrary function: 
\begin{equation}
S=\mu\int d\lambda \,E\, f\left(\nabla y^i\nabla y^i\right)=
\int dt\sqrt{\dot{x}^\mu \dot{x}_\mu} (\pd\chi)F\left[
\frac{(\dot{x}^\mu \pd_\mu\chi)^2}{(\pd\chi)^2\dot{x}^\mu \dot{x}_\mu}\right].
\end{equation}
This is precisely the action used in \cite{Cuomo}.

\subsection{3+1 dimensions}

From \eqref{eqAppSBvortexGauged}, the coset is written as
\begin{equation}
\Omega=e^{i y^a P_a}e^{i\sigma D}e^{i\eta^i J_{0i}}e^{i\xi^n J_{n3}}e^{i\chi Q}.
\end{equation}
We use indices $m,n=1,2$ and $p,q=0,3$. One can compute the MC one form as before
\begin{equation}
\Omega^{-1}D_\mu \Omega
=i E^a_\mu\left(P_a+\nabla_a\sigma D+\nabla_a\chi Q+\nabla_a\eta^i
J_{0i}+\nabla_a\xi^n J_{n3}
+\Omega^{12}_a J_{12}
\right);
\end{equation}
with
\begin{equation}
E^a_\mu=e^{-\sigma}e^c_\mu\Lambda_c^{\,b}R_b^{\,a},\qquad
\nabla_a\chi=e^{\sigma}e^{\mu}_c\Lambda^c_{\,b}R^b_{\,a}\pd_\mu\chi,\qquad
\nabla_a\sigma=e^{\sigma}e^{\mu}_c\Lambda^c_{\,b}R^b_{\,a}\left(\pd_\mu\sigma+A_\mu\right).
\end{equation}
Here we introduced another Lorentz matrix $(e^{-i\xi^n J_{n3}})_{\;b}^{a}=R_b^{\;a}$.
Curvature invariants are written as before. The MC form projected on the vortex world-sheet $X^\mu(\tau,\sigma)$ reads:
\begin{equation}
\pd_\alpha X^\mu \Omega^{-1}D_\mu \Omega
=i E_\alpha^p \left(P_p+\nabla_p y^n P_n+\nabla_p\sigma D+\nabla_p\chi Q
+\nabla_p\eta^i
J_{0i}
+\nabla_p\xi^n J_{n3}+
 \Omega_p^{12} J_{12}
\right),
\end{equation}
where $\alpha=\tau,\sigma$ and
\begin{equation}
\begin{gathered}
E_\alpha^p=\pd_\alpha X^\mu e^{-\sigma}e^c_\mu\Lambda_c^{\,b}R_b^{\,p},\qquad
\nabla_p y^n=E^\alpha_p\pd_\alpha X^\mu 
 e^{-\sigma}e^c_\mu\Lambda_c^{\,b}R_b^{\,n},\\
\nabla_p\chi=E^\alpha_p\pd_\alpha X^\mu\pd_\mu\chi,
\qquad
 \nabla_p\sigma=E^\alpha_p\pd_\alpha X^\mu \left(\pd_\mu\sigma+A_\mu\right).
 \end{gathered}
\end{equation}
Here $E^\alpha_p$ is the inverse of the world-sheet vierbein: $E_\alpha^p E^\alpha_q=\delta^p_q$, $E_\alpha^p E^\beta_p=\delta^\alpha_\beta$.
As before, the IHCs \eqref{eqAppIHC1} are imposed. Since $[P_3,J_{n3}]\sim P_n$, we can also eliminate $\xi^n$ imposing the following IHC 
\begin{equation}
\nabla_3 y^n=0\qquad\implies \qquad 
\frac{\xi^n}{\xi}\tan\xi=\frac{v^n}{v^3},
\end{equation}
where the vector $v^i$ is given by
\begin{equation}
v^i=\frac{\left(\pd_3 X^\mu \pd_\mu\chi\right)
\left(\pd_0 X^\mu e_\mu^c\Lambda_c^{\,i}\right)-\left(\pd_0 X^\mu \pd_\mu\chi\right)
\left(\pd_3 X^\mu e_\mu^c\Lambda_c^{\; i}\right)
}{(\pd\chi)\sqrt{-\text{det}(G_{\alpha\beta})\,h_{\alpha\beta}G^{\alpha\beta}}}.
\end{equation}
Here $G_{\alpha\beta}$ and $h_{\alpha\beta}$ are:
\begin{equation}
G_{\alpha\beta}=g_{\mu\nu}\pd_\alpha X^\mu\pd_\beta X^\nu,\qquad
h_{\alpha\beta}=\frac{\pd_\mu\chi\pd_\nu\chi}{(\pd\chi)^2}\pd_\alpha X^\mu\pd_\beta X^\nu.
\end{equation}
These expression agree with the previous definitions \eqref{eqStringMetricG} and \eqref{eqStringMetrich}.
Since $\nabla_p\chi=\mu\delta_p^0$ and
$\nabla_p\sigma=0$, leading order invariants are built out of the following objects
\begin{equation}\label{eqAppdetE}
\mu^2\text{det}(E^p_\alpha)= (\pd\chi)^2\sqrt{|\text{det}(G_{\alpha\beta})|}\,
\sqrt{G^{\alpha\beta}h_{\alpha\beta}},\qquad
\nabla_0 y^n\nabla_0 y^n=1-\frac{1}{h_{\alpha\beta}G^{\alpha\beta}}.
\end{equation}
One finally writes the leading order action as
\begin{equation}
S=\mu^2\int d\tau d\sigma\det{E^p_\alpha} \,
f\left(\nabla_0 y^n\nabla_0 y^n\right)=
\int d\tau d\sigma(\pd\chi)^2\sqrt{|\text{det}(G_{\alpha\beta})|}F[G^{\alpha\beta} h_{\alpha\beta}].
\end{equation}
Using \eqref{eqCurrent}, this agrees with the last line in \eqref{eqAction1}.

\bibliography{Biblio}
	\bibliographystyle{JHEP.bst}

\end{document}